\documentclass[reprint,twocolumn,prb,amsmath,amssymb,aps]{revtex4-2} 

\usepackage{graphicx}  
\usepackage{dcolumn}   
\usepackage{bm}        
\usepackage{hyperref}  
\hypersetup{colorlinks=true, linkcolor=blue!95!black!85!yellow, citecolor=blue!95!black!85!yellow, urlcolor=blue!95!black!85!yellow} 
\usepackage{physics}   
\usepackage{newtxtext} 
\usepackage[cmintegrals]{newtxmath} 
\usepackage{booktabs}  
\usepackage{multirow}  
\usepackage{xcolor}    
\usepackage{float}     

\usepackage{tikz}
\newcommand{\orcidicon}{%
    \begin{tikzpicture}
        \draw[lime, fill=lime] (0,0) circle [radius=0.16] node[white] {{\fontfamily{qag}\selectfont \tiny ID}};
        \draw[white, fill=white] (-0.0625,0.006) circle [radius=0.007];
    \end{tikzpicture}
    \hspace{-2.9mm}
}
\foreach \x in {A, B, C}{\expandafter\xdef\csname orcid\x\endcsname{\noexpand\href{https://orcid.org/\csname orcidauthor\x\endcsname}{\noexpand\orcidicon}}}

\newcommand{\iitm}{School of Mechanical and Materials Engineering, Indian Institute of Technology Mandi, Kamand 175075, India}

\begin{document}

\title{Electronic structure, excitonic optical response, and defect stability of ${\rm RbTlCl_3}$ from first-principles}

\author{Siddharth \orcidA{}}
\email{sid1391998@gmail.com}
\author{Vinod Kumar Solet \orcidB{}}
\author{Sudhir K. Pandey \orcidC{}}
\email{sudhir@iitmandi.ac.in}
\affiliation{\iitm}
\date{\today}

\begin{abstract}
Here, we present a detailed many-body \textit{ab initio} study of the RbTlCl$_{3}$ perovskite. The electronic and optical properties, both with and without spin-orbit coupling, have been calculated using density functional theory (DFT) and many-body excited-state calculations. The band gap, which is indirect in nature, are found to be of 0.94 and 0.95 eV from PBE-GGA exchange correlation functional with and without spin-orbit coupling, respectively. The optical properties have been computed using methods based on independent particle approximation (IPA), IPA with scissor correction (IQPA), local-field effects (LFEs) at random phase approximation (RPA) level, and the Bethe-Salpeter equation (BSE). The estimated highest value of the imaginary part of the dielectric function using IQPA is $\sim$7 at 2 eV, which slightly decreases to $\sim$5.7 due to LFEs. Within the BSE, the peak value is obtained to be maximum at 1.6 eV with a magnitude of $\sim$10.8, which indicates the strong excitonic effect below the scissor corrected direct gap. A large number of bright and dark bound excitons are found, where the binding energies of four main bound bright excitons are found to be in the range of 315-365 meV. The main bound bright exciton is localized (delocalized) in the reciprocal (real) space. Also, the spectroscopic limited maximum photovoltaic efficiency is estimated to be 15.4 \% for a 0.5 $\mu$m thin film. Finally, the positive phonon frequencies and thermodynamical stabile region obtained for this perovskite indicate its dynamical and thermodynamical stability. Using thermodynamic formalism of chemical potentials, $V_{Cl}$ ($V_{Tl}$) is identified as the most stable neutral point defect in Cl-poor (rich) environment. These findings reveal a significant excitonic effect in RbTlCl$_{3}$ and provide valuable theoretical insights for discovering new potential photovoltaic materials within the inorganic perovskite family.
    
\end{abstract}

\maketitle

\section{Introduction} 
\setlength{\parindent}{3em}
Perovskite semiconducting materials play an important role in the making of photovoltaic (PV) solar cell technologies \cite{huang2021perovskite, yang2017iodide}. Recently, perovskite solar cells have gained significant attention due to their rapidly improving power conversion efficiency (PCE) \cite{xiong2024reducing, kojima2009organometal, basera2022chalcogenide, palummo2020halide, NREL}. Moreover, these materials have good optical properties, such as a high absorption coefficient \cite{de2014organometallic} and favorable band gaps \cite{kulkarni2014band}, which make them suitable for PV applications. Therefore, researchers worldwide are actively exploring perovskite compounds to achieve maximum efficiency in PV devices.

At the computational level, a material's band gap and optical absorption characteristics are the primary factors that determine its potential for PV applications. The electronic and optical properties of various semiconductors have been thoroughly examined in recent years using first-principles-based density functional theory (DFT) \cite{han2022ground, nabi2024lead, basera2020reducing, du2021cerium, solet2025many, solet2024mg}.  But the DFT method is only accurate for predicting ground-state properties and provides only an approximate description of excited-state phenomena, such as band gaps and optical properties \cite{hybertsen1985first, onida2002electronic}. Because single-particle DFT calculates properties within the independent particle approximation (IPA) \cite{bechstedt2016many}, the optical spectra obtained at this level generally differ from experimental results \cite{rohlfing2000electron, benedict1998optical}. Hybrid functionals \cite{hinuma2017band, muscat2001prediction, heyd2005energy} or GW methods \cite{solet2025many, onida2002electronic} overcome these limitations and provide a more realistic description of the band gap. In light of this, more advanced theoretical approaches beyond single-particle DFT are required to evaluate PV materials accurately. When a photon is absorbed, it excites an electron from the valence band (VB) to the conduction band (CB), leaving behind a hole in the VB, forming a bound electron-hole (e-h) pair called an exciton. The excitonic information can not be achieved within IPA method \cite{rohlfing2000electron}. To include these effects, many-body perturbation theory based on the Bethe-Salpeter equation (BSE) has been used  \cite{onida2002electronic}. The absorption spectra produced using the BSE method are closely align with the experimental results for many materials \cite{solet2024mg, pulci2001many}. 

While accurate calculation of optical properties is critical, it is equally important to assess how these properties translate into actual device performance, which is typically measured using the PCE. The device performance is impacted by factors such as the absorption spectra and the band gap \cite{yu2012identification, kirchartz2018makes}. Traditionally, the efficiency of single-junction solar absorbers had been calculated using the Shockley-Queisser (SQ) limit \cite{shockley1961detailed}. This method relies on the thermodynamic principle of detailed balance, using the material's band gap with a absorption as a step function and only looking at radiative recombination. Researchers use an advanced version of the SQ limit, known as the spectroscopic limited maximum efficiency (SLME) model \cite{yu2012identification}, for real materials. The SLME calculation utilizes the actual absorption spectra, the optical and fundamental band gaps, and the thickness of the absorbing layer \cite{yu2012identification}.

Additionally, native point defects play an important role in determining the electronic and optical properties of semiconducting materials \cite{rahman2025defect, chatratin2019role}. It has been seen that native defects often control e-h recombination processes, mainly in applications such as optoelectronics \cite{solet2024mg}. In many cases, these defects act as non-radiative recombination centers, where electrons and holes recombine without emitting photons, which causes energy loss through lattice vibrations. This process affects the performance of PV devices \cite{bang2016carrier}. On the other hand, carrier transport properties can be tuned by controlling point defects and the resulting carrier concentrations, as seen in thermoelectric materials \cite{snyder2008complex}. The formation and concentration of these native defects are influenced by the thermodynamic stability conditions of the host material, which are governed by the chemical potentials of its constituent elements. In different growth conditions, the allowed range of chemical potentials defines the stability region of the compound \cite{buckeridge2014automated}. This is also useful for finding favorable defects and their formation energies. Therefore, understanding the interplay between native defects and thermodynamic stability is essential for tailoring material properties for specific applications.

Finding materials suitable for PV applications poses a significant challenge. For that, researchers have explored various classes of materials, including skutterudites \cite{chen2003recent}, Heusler compounds \cite{chen2003recent, solet2022first, solet2023ab}, complex chalcogenides \cite{khare2020electronic}, perovskites \cite{huang2023inorganic}, semiconducting silicides \cite{solet2024mg, sharma2014first}, etc., for their potential use in energy production. Among them, perovskite materials have gained considerable attention for their PV application \cite{zhou2014interface}. Generally, perovskite solar cells have majorly used hybrid (organic-inorganic) halide perovskite materials \cite{berhe2016organometal, baikie2013synthesis, d2014excitons, ganose2017beyond}. These materials have ABX$_{3}$ type of structure in which A is an organic monovalent cation, B is an inorganic divalent cation, and X is a halogen anion \cite{glazer1975simple}. Hybrid perovskite semiconducting materials have many good characteristics for PV applications. But the main issue is the organic cation in these compounds. In these types of hybrid perovskites, the organic part is sensitive to moisture and oxygen, which leads to the degradation \cite{berhe2016organometal}. To address this issue, it has been proposed to substitute the organic cation with inorganic monovalent alkali cation like rubidium, as they exhibit lower moisture sensitivity and gain a more stable perovskite structure \cite{jong2018first, li2017all, saliba2016incorporation}. Following that, various perovskites have been predicted to be semiconductors at the theoretical \cite{endres2016valence, nyayban2020first, labrim2023optoelectronic, sebastian2015excitonic} as well as experimental levels \cite{sebastian2015excitonic, stoumpos2015renaissance, wells1893casium}. In addition, alkali halide-based perovskites have attracted significant interest in the research community due to their suitable PV properties \cite{sarhani2023ab, behera2024first, sarhani2023ab, maskar2024structural}.

\begin{figure} 
	\label{fig:RbTlCl3_structure}
	\includegraphics[width=8cm, height=6cm]{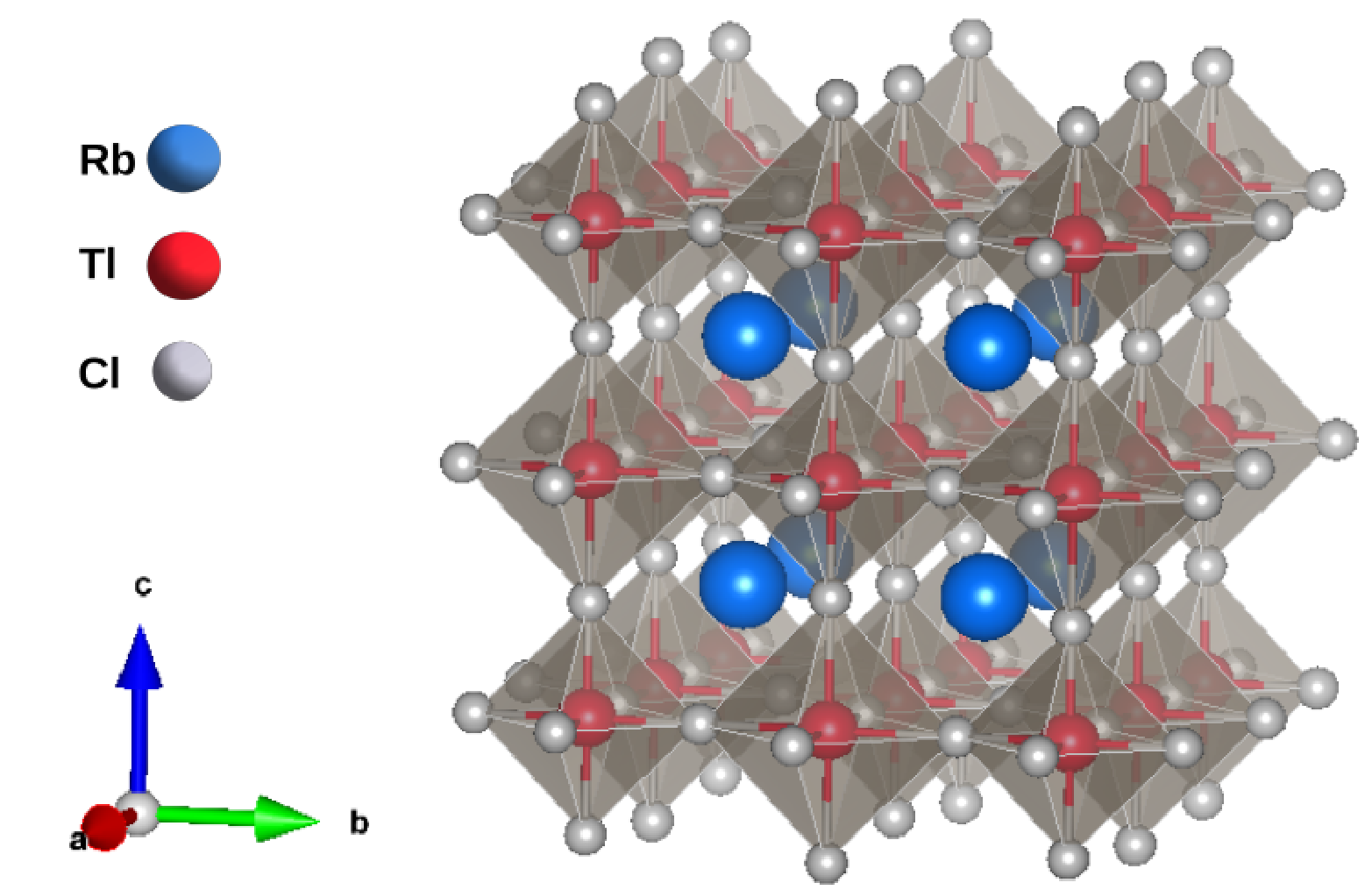}
	\caption{Crystal structure of RbTlCl$_{3}$}
\end{figure}

In this work, we focus our attention on a specific member of the inorganic perovskite family that exhibits several characteristics desirable for potential PV applications. From a literature survey, we identified a face-centered cubic RbTlCl$_{3}$ compound in the series of alkali-based perovskites worthy of investigation for potential PV applications. One of the reasons behind selecting this perovskite is to have a semiconducting band gap of $\sim$0.74 eV \cite{hase2017evolution}. For the prediction of perovskite structure, the Goldschmidt tolerance factor/ideality factor has to be calculated. For the ideal cubic perovskite structure, the ideality factor varies from $\sim$0.9 to 1 \cite{li2016stabilizing}. This requirement is met by RbTlCl$_{3}$, which has an ideality factor of about $\sim$0.9, making it more stable than other hybrid perovskites. The cubic perovskite crystal structure of RbTlCl$_{3}$ is shown in Fig.~\ref{fig:RbTlCl3_structure} \cite{momma2008vesta}. RbTlCl$_{3}$ has a similar structure as BaBiO$_{3}$ and CsTlCl/F$_{3}$ \cite{yin2013rational}; these compounds have the B site as mixed valent cations such as Bi (Bi$^{+3}$ and Bi$^{+5}$) and Tl (Tl$^{+1}$ and Tl$^{+3}$) \cite{saparov2016organic}. To date, a comprehensive analysis of RbTlCl$_{3}$ as a PV material has not been carried out at either the experimental or theoretical level. To evaluate the potential of RbTlCl$_{3}$ for PV applications, we have performed a detailed first-principles investigation of its electronic structure, excitonic optical response, and defect stability.

The remainder of this paper is organized as follows. In Section~\ref{sec:tm}, the theoretical framework and computational details employed in this work are presented. Section III contains the results and discussion. In Section~\ref{sec:band}, the electronic properties are discussed through the band structure and total and partial density of states (TDOS and PDOS). The optical properties are investigated in Section~\ref{sec:optical} using DFT and many-body excited-state calculations, with and without spin-orbit coupling (SOC), within different approximations including IPA, independent quasi-particle (QP) approximation (IQPA), random phase approximation (RPA), and BSE. In Section~\ref{sec:exciton}, the excitonic properties are analyzed in both momentum and real space. In Section~\ref{sec:SLME} presents the SLME calculated within different approximations. Finally, in Section~\ref{sec:stability}, the dynamical and thermodynamic stabilities are analyzed. The thermodynamic stability is used to determine the range of chemical potentials under which RbTlCl$_3$ can exist, and intrinsic neutral point defects along with their formation energies are examined to understand the defect behavior of the material.

\section{\label{sec:tm}THEORY AND METHODS}
\subsection{Optical abosrption}
One of the main objectives of this work is to obtain accurate optical properties by considering various theoretical approximations—IPA, IQPA, RPA, and BSE. In the IPA, the optical response is calculated directly from the Kohn-Sham electronic structure without the inclusion of QP corrections and e-h interactions. It is important to note that the IQPA builds upon the IPA by introducing QP eigenvalues, which typically derive from a $GW$ calculation. However, to efficiently simulate these QP corrections here, we adopt a scissor operator ($\Delta$) to correct the direct band gap. This rigidly shifts the conduction band energies, aligning the direct gap of the DFT calculation with the mBJ value. Thus, in the present work, the IQPA spectra correspond to the IPA eigenvalues adjusted via this scissor correction. The local-field effects arising from the inhomogeneous electron density are subsequently treated at the RPA level through the inversion of the microscopic dielectric function \cite{onida2002electronic, gulans2014exciting}. In this approach, these local fields modify the effective electric field inside the crystal and therefore the dielectric response. The BSE includes the screened e-h interaction through the excitonic Hamiltonian, enabling an accurate description of excitonic effects and optical absorption beyond the independent-particle picture \cite{onida2002electronic, solet2024mg}.

The BSE can be written as an eigenvalue problem of an effective two-QP excitonic Hamiltonian ($H^{\mathrm{BSE}}$). This Hamiltonian is solved here within the Tamm-Dancoff approximation (TDA), which neglects the coupling between the resonant (excitation) and anti-resonant (de-excitation) terms \cite{vorwerk2019bethe, gulans2014exciting}. 
\begin{eqnarray}
	\sum_{\nu' c' \mathbf{k'}} H^{BSE}_{\nu c \mathbf{k}, \nu' c' \mathbf{k'}} A^{\lambda}_{\nu' c' \mathbf{k'}} = E^{\lambda}A^{\lambda}_{\nu c \mathbf{k}}
\end{eqnarray}
Where $E^{\lambda}$ and $A^{\lambda}$ are the eigenvalue and eigenfunction of exciton, respectively. The two-particle BSE equation can be written as \cite{chang2000excitons}:  
\begin{equation}
	\begin{split}
		\left(E_{c\mathbf{k}} - E_{\nu \mathbf{k}} + \Delta\right) A^{\lambda}_{\nu c\mathbf{k}} 
		+ \sum_{\nu' c' \mathbf{k'}} 
		K^{\nu' c' \mathbf{k'}}_{\nu c \mathbf{k}} A^{\lambda}_{\nu' c'\mathbf{k'}}
		= E^{\lambda} A^{\lambda}_{\nu c\mathbf{k}}
	\end{split}
\end{equation}
Here, the first term on the left-hand side represents the diagonal part, where $\nu$ and $c$ denote the occupied and unoccupied states, respectively, and $\Delta$ is the scissor operator. The second term, $K^{\nu' c' \mathbf{k'}}_{\nu c \mathbf{k}}$, in Eq.~(2), corresponds to the e-h interaction kernel. This kernel can be expressed as $K = 2K_{x} + K_{d}$, where $K_{x}$ and $K_{d}$ are the exchange and direct Coulomb interactions between the electron and hole, respectively \cite{rohlfing1998electron}. In the IQPA approach, only the diagonal term is retained, i.e., $K=0$. In the RPA, the $K_x$ is additionally included. For spin-singlet states within the BSE formalism, the $K_d$ is further included. Thus, the full BSE Hamiltonian in Eq.~(2) explicitly accounts for e-h interactions and captures the excitonic effects in the system.

In optical absorption spectroscopy, the photon's wave vector is negligible relative to the dimensions of the crystal lattice. Consequently, in the long-wavelength limit as ${q\to 0}$, the $\epsilon_{2}(\omega)$ is given by \cite{onida2002electronic, solet2025many}:
\begin{equation}
	\begin{split}
		\epsilon_{2}(\omega) = \lim_{q \to 0} \, \frac{8\pi^{2}}{Vq^{2}} \, \sum_{\lambda} \, \left|\sum_{\nu c k} A^{\lambda}_{\nu ck}\langle\nu \textbf{k}|e^{-i\textbf{q}.\textbf{r}}|c\textbf{k} \rangle \right|^2 \\ \times\delta (E^{\lambda}-\omega)
	\end{split}
\end{equation}
Where V represents the volume of a crystal. The part inside the modulus square has two components: one is the excitonic state ($A^{\lambda}_{\nu ck}$), and the other is the momentum matrix element $\langle\nu \textbf{k}|e^{-i\textbf{q}.\textbf{r}}|c\textbf{k} \rangle$. This whole modulus squared term is called the oscillator strength of the $\lambda^{th}$ exciton. The delta function $\delta (E^{\lambda}-\omega)$ ensures the energy conservation during absorption process. In IPA, the Eq. 3 can be modified as:
\begin{equation}
	\begin{split}
		\epsilon_{2}(\omega) = \lim_{q \to 0} \, \frac{8\pi^{2}}{Vq^{2}} \, \left|\sum_{\nu c k} \langle\nu \textbf{k}|e^{-i\textbf{q}.\textbf{r}}|c\textbf{k} \rangle \right|^2 \\ \times\delta(\varepsilon_{c,\mathbf{k}} - \varepsilon_{\nu,\mathbf{k}} - \omega)
	\end{split}
\end{equation}

The real part of the macroscopic dielectric function $\epsilon_{1}(\omega)$ can be calculated by the Kramers-Kronig relation \cite{dressel2002electrodynamics}, which is given by: 
\begin{eqnarray}
	\epsilon_{1}(\omega) = 1+\frac{2P}{\pi}\int_{0}^{\infty}\frac{\omega'\epsilon_{2}(\omega')}{(\omega'^{2}-\omega^{2})}d\omega' 
\end{eqnarray}
Where P is the Cauchy principal value of the integral and $\omega'$ is a variable used for integration across the full range of frequencies. All the remaining PV properties, such as the real $\sigma_{1}(\omega)$ and imaginary $\sigma_{2}(\omega)$ parts of optical conductivity $\sigma(\omega)$, complex refractive index $\tilde{n}(\omega)$, absorption coefficient $\alpha(\omega)$, and reflectivity r($\omega$), can be easily computed from the dielectric function \cite{ambrosch2006linear}.

The spatial spread of excitons in momentum and real space can be analyzed after solving the BSE. In reciprocal space, the distribution of the excitonic wave function is associated with $A_{vck}^{\lambda}$. The weights associated with transitions between the VB and CB at each $\mathbf{k}$ point can be defined as \cite{bechstedt2016many}:
\begin{subequations}\label{eq:weights}
	\begin{align}
		w^{\lambda}_{v\mathbf{k}} &= \sum_{c} \left| A^{\lambda}_{v c \mathbf{k}} \right|^2, \\
		w^{\lambda}_{c\mathbf{k}} &= \sum_{v} \left| A^{\lambda}_{v c \mathbf{k}} \right|^2.
	\end{align}
\end{subequations}
Also, the real-space excitonic wave function $\Psi_{\lambda}(\textbf{r}_{e},\textbf{r}_{h})$ consists of six coordinates, with three corresponding to the electron and three to the hole. It describes the probability amplitude of finding the electron at position $\textbf{r}_{e}$ when the hole is kept fixed at $\textbf{r}_{h}$ for convenience. It is constructed as a superposition of products of Bloch functions from filled ($\phi_{\nu \textbf{k}}(\textbf{r}_{h})$) and empty ($\phi_{c\textbf{k}}(\textbf{r}_{e})$) states, weighted by $A_{\nu c\textbf{k}}^{\lambda}$ \cite{bechstedt2016many}:
\begin{equation}
	\Psi_{\lambda}(\textbf{r}_{e},\textbf{r}_{h}) = \sum_{\nu c\textbf{k}} A_{vck}^{\lambda} \phi_{c\textbf{k}}(\textbf{r}_{e}) \phi_{\nu \textbf{k}}(\textbf{r}_{h})
\end{equation}
\subsection{Spectroscopic limited maximum efficiency (SLME)}
For the evaluation of the PCE of the material, SLME is employed. The PCE is calculated as the ratio of the maximum output power density ($P_o$) to the total incident solar power density ($P_i$):
\begin{eqnarray}
	SLME = \frac{P_o}{P_i}
\end{eqnarray}
The P$_{o}$ has been obtained by numerically maximizing the product of current density (J) and voltage (V). For a solar cell exposed to a photon flux I$_{sun}$ at temperature T, and assuming it behaves like an ideal diode \cite{green2006third}, the relationship between $J$ and $V$ is given by \cite{kirchartz2018makes}:
\begin{eqnarray}
	J = J_{sc} - J_{0}(1 - e^{\frac{eV}{k_{B}T}})
\end{eqnarray}
Where k$_{B}$ is the Boltzmann constant. This equation describes how $J$ changes with $V$, accounting for both the light-generated current and material’s current in the dark. The short-circuit current density, J$_{sc}$, is defined as:
\begin{eqnarray}
	J_{sc} = e\int_{0}^{\infty}a(E)I_{sun}(E)dE
\end{eqnarray}
In this equation, $e$ is the elementary charge and $a(E)$ is the photon absorptivity. The $J_{0}$ in second term in Eq. (9), represents the reverse saturation current at equilibrium in the dark, which has been considered as the sum of radiative ($J_{0}^{r}$) and non-radiative ($J_{0}^{nr}$) $e$–$h$ recombination currents. The total $J_{0}$ is given by $J_{0} = J_{0}^{r} + J_{0}^{nr} = J_{0}^{r}/f_{r}$.
In this context, $f_{r}$ represents the fraction of radiative recombination current. In SLME, $f_{r}$ is approximated as $e^{-{\varDelta/k_{B}T}}$, where $\varDelta = E^{da}_{g} - E_{g}$, with $E^{da}_{g}$ being the direct allowed gap and $E_{g}$ the fundamental gap. According to the principle of detailed balance, a solar cell must emit and absorb light at the same rate through its surfaces when in the dark and at equilibrium. Therefore, $J_{0}^{r}$ has been found by calculating how many black-body photons from the surroundings are absorbed through the front surface \cite{tiedje1984limiting}. We can calculate it as follows:
\begin{eqnarray}
	J_{0}^{r} = e\pi\int_{0}^{\infty}a(E)I_{bb}(E,T)dE
\end{eqnarray}
where $I_{bb}(E,T)$ is the black-body photon flux at temperature $T$. The absorptivity $a(E)$ is expressed as $(1 - e^{-2\alpha(E)L})$, where $\alpha(E)$ and $L$ are the absorption coefficient and the thickness of the material, respectively \cite{tiedje1984limiting}.

\subsection{\label{sec:diff}Native defects} 
To characterize point defects in a system, an important quantity is the defect formation energy. For a point defect $d$ in charge state $q$, the formation energy within the supercell approach is defined as \cite{freysoldt2014first, solet2024mg, liang2021first}:
\begin{equation}
	\Delta E_f(d^q) = E(d) - E(\mathrm{bulk}) - \sum_i n_i \mu_i
	+ q(\varepsilon_v + \varepsilon_F),
\end{equation}
where $E(d)$ and $E(\mathrm{bulk})$ are the calculated energies of the defective and non-defective (host) supercells, respectively; the $N_i$ represents the number of defect atoms $i$ added to ($n_i > 0$) or removed from ($n_i < 0$) the host supercell to form the defect; $\mu_i$ are the corresponding chemical potentials obtained from the DFT computed thermodynamic stable region under different growth conditions \cite{buckeridge2014automated}, and the final term accounts for the exchange of electrons with the reservoir at Fermi level $\varepsilon_F$, referenced to the VB maximum (VBM), $\varepsilon_v$. In the present study, finite-size correction term \cite{freysoldt2014first} associated with the supercell approach is neglected. All defect formation energy calculations presented in this work are restricted to neutral charge states ($q=0$).

\subsection{Computational details}
All DFT and beyond DFT calculations have been calculated with the help of all-electron full-potential $exciting$ code \cite{gulans2014exciting}. The (linearized) augmented plane-wave plus local orbital basis set has been used to solve the Kohn-Sham (KS) equation. The Perdew, Burke, and Ernzerhof-based generalized gradient approximation (PBE-GGA) \cite{GGA-PBE} exchange and correlation (XC) functional has been used throughout the calculations. The ground state calculation for SOC and the non-SOC case has been done over the \textbf{k}-point mesh of 10$\times$10$\times$10 with an energy convergence limit of 1$\times$$10^{-6}$ Hartree and a plane wave cutoff of R$_{MT}$$\mid\mathbf{G + K}\mid_{max}$ = 7.0. The phonon calculation is performed within the framework of density-functional perturbation theory (DFPT) under PBE-GGA functional with $4\times4\times4$ $\mathbf{k}$-point and $2\times2\times2$ $\mathbf{q}$-point grids. In the case of SOC, 20 unoccupied states are considered for the solution to the electronic ground state. For all calculations, the optimized lattice parameter of $\sim$11.27~$\mathrm{\AA}$ \cite{SM} has been used within the space group of Fm$\bar{3}$m (space group No.~225). In the present study, structural optimization is performed using the PBE-GGA XC functional, which is also employed for the calculations reported in the Materials Project database \cite{horton2025accelerated, jain2011formation}, where the calculations are carried out using the VASP code \cite{kresse1996efficient, kresse1999ultrasoft}. Details of the volume and atomic position optimizations are provided in the Supplementary Material (SM) \cite{SM}. The muffin-tin radii ($R_{MT}$) of all the constituent atoms (except Cl) are set to be 2.5 Bohr, while for Cl, it is 2.23 Bohr.

The thermodynamic stability is evaluated using the \textsc{CPLAP} code by comparing the free energy of formation of the target compound with those of all relevant competing phases, using the species, stoichiometry, and free energies of formation as inputs \cite{buckeridge2014automated}. The initial theoretical structural information for all competing phases is obtained from the Materials Project database \cite{horton2025accelerated}. The optimized lattice parameters of the competing phases have been used for the thermodynamic stability calculations. Details of the volume optimization have been provided in the SM \cite{SM}. Further, the $\mathbf{k}$-point meshes for RbCl ($Pm\bar{3}m$), TlCl ($Pm\bar{3}m$), and Rb$_3$TlCl$_6$ ($Fm\bar{3}m$) are chosen as $10\times10\times10$ for the total-energy DFT calculations, while an $8\times5\times4$ mesh is used for TlCl$_3$ (Pnma). For the elemental reference phases, $\mathbf{k}$-point meshes of $20\times20\times20$ and $18\times18\times12$ are used for Rb ($Im\bar{3}m$) \cite{barrett1956x} and Tl ($P6_3/mmc$) \cite{horton2025accelerated}, respectively.  To obtain the total energy of Cl, we have performed calculation on Cl$_2$ molecule as implemented in $exciting$ code \cite{powell1984temperature}.

In this study, the optical properties have been obtained at various levels of approximations, which are IPA, IQPA, RPA, and BSE. We have calculated single-particle eigenstates and eigenenergies for many body calculations via a non-self-consistent field calculation using a \textbf{k}-mesh grid of 4$\times$4$\times$4 with a shift of (0.097, 0.273, 0.493), along with 50 unoccupied states that have been used in all approximations. To simulate the self-energy effect, we apply the scissor correction on the KS eigenvalues. Note that the scissor operator value of $\sim$459 (465) meV for the non-SOC (SOC) case has been used, which is the difference between the direct band gap obtained using the modified Becke-Johnson potential \cite{tran2009accurate} (within the $WIEN2k$ code \cite{blaha2020wien2k}) and the PBE-GGA functional. Furthermore, the \textbf{q}-mesh of a 4$\times$4$\times$4 with 100 empty states is used for screening calculation. It is important to note that the RPA implemented in $exciting$ code inherently includes the LFEs. In the present calculations, a threshold energy of 3.0 Hartree has been utilized for the LFEs in the RPA calculation. In addition, the BSE Hamiltonian is built in the transition space using the top 15 (29) occupied bands and the first 20 unoccupied bands for the non-SOC (SOC) calculations. The Lorentzian broadening is taken at $\sim$0.1 eV with no intraband contributions added.

\section{Results and discussion} \label{sec:results}
\subsection{\label{sec:band}Electronic properties}
The electronic band structure plays a crucial role in determining the applicability of PV materials in solar technology. Since optical processes are associated with direct band-to-band transitions, the number of excited electrons in unoccupied states strongly depends on the nature and type of the band gap in PV materials, as also reflected in Eq. 3. The electronic dispersion of RbTlCl$_{3}$ is calculated with PBE-GGA XC functional with and without SOC. 

\begin{figure}[ht]
	\label{fig:band}
	\includegraphics[width=8.5cm, height=6cm]{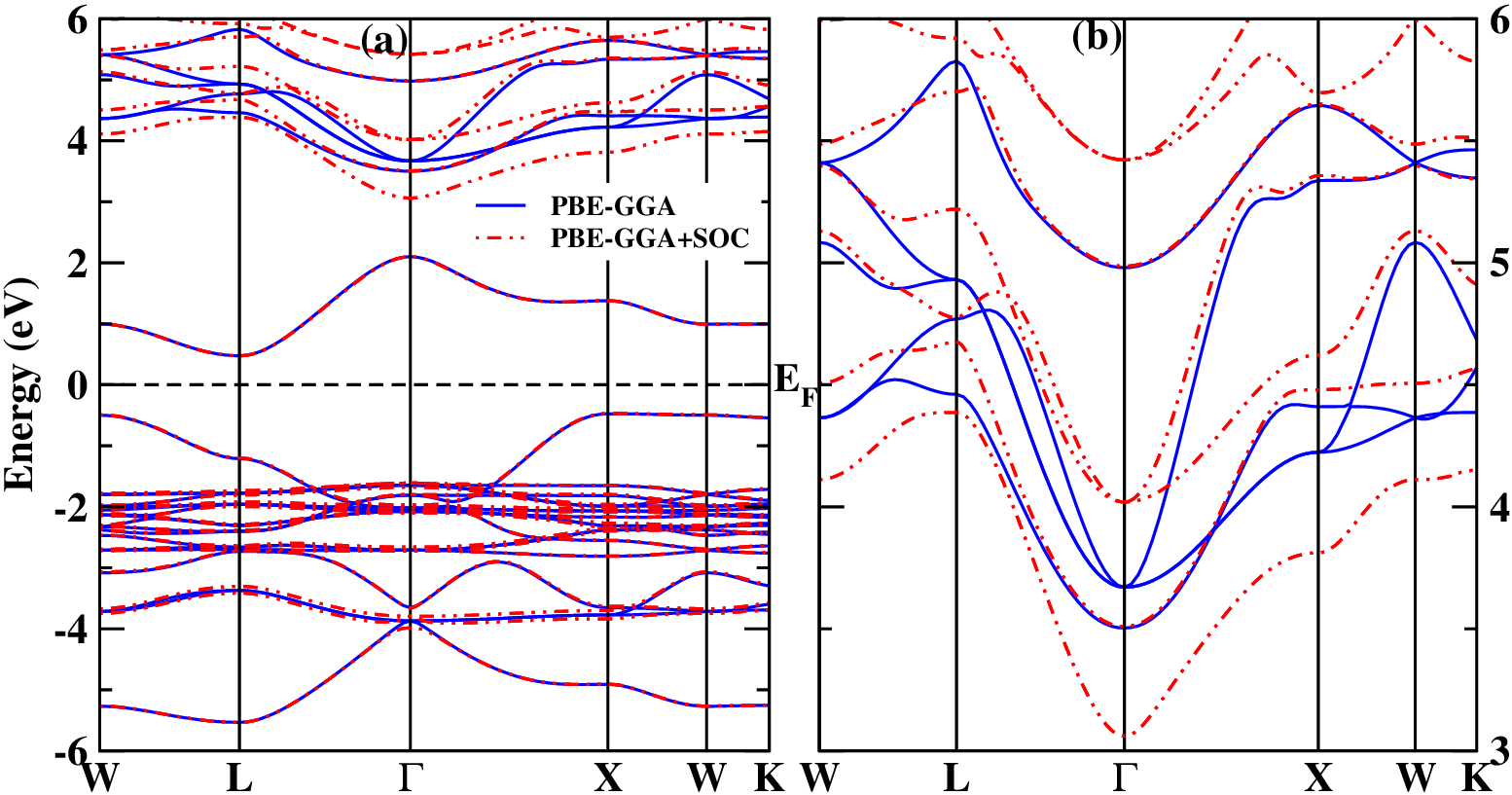} 
	\caption{(a) Electronic dispersion of RbTlCl$_{3}$ in the high symmetric direction obtained from PBE-GGA with and without SOC and (b) magnified 3 to 6 eV region of band structure within SOC} 
\end{figure} 
The dispersion curve of RbTlCl$_{3}$ in Fig. 2(a) is analyzed within the first Brillouin zone along the high symmetric k-path direction $W$-$L$-$\Gamma $-$X$-$W$-$K$. The horizontal dashed line in the middle of the band gap represents the Fermi energy ($E_{F}$). Bands from both functionals have shown almost similar characteristics. The VBM is present at the high symmetric point $X$, and the CB minimum (CBM) is present at the high symmetric point $L$, which makes it evident that RbTlCl$_{3}$ is an indirect band gap semiconducting compound. The indirect band gaps of RbTlCl$_{3}$ obtained from PBE-GGA in the presence and absence of SOC are $\sim$0.94 eV and $\sim$0.95 eV, respectively. It is well known that direct transitions play a vital role in the optical properties of compounds \cite{fox2002optical}. The direct Kohn-Sham band gap in this compound with and without SOC are found at the $W$ high symmetric point, with values of $\sim$1.49 and $\sim$1.50 eV, respectively.

In a previous study on metal halide perovskites \cite{hussain2021spin}, it has been observed that SOC lifts the degeneracy of bands and modifies the band gap. Therefore, it is crucial to investigate the effect of SOC on the band structure, as it is essential for this compound due to the presence of the heavy Tl element. Consequently, we apply SOC on top of PBE-GGA (PBE-GGA+SOC) for this compound to see the effect of SOC on the band structure. The effect of SOC on the band gaps (fundamental and direct) is negligible in this compound. As a result, the direct gap decreases slightly (by approximately 10 meV), which is not significant enough to affect the optical properties of the material. However, the high and low energy regions (above 3 eV and below -3 eV) are affected within SOC. Due to this, we can observe the effect of SOC on optical properties in the high-energy region. For the more detailed analysis of these effects on RbTlCl$_{3}$, Fig. 2(b) highlights the band structure specifically within the energy range of $\sim$3-6 eV. It can be observed that the threefold-degenerate bands at the $\Gamma$ point split due to SOC into twofold-degenerate bands at higher energy and a single band at lower energy. A similar change is also observed at the $W$ point, where two degenerate bands become non-degenerate in the presence of SOC. Unfortunately, no experimental information is available to compare the electronic properties of RbTlCl$_{3}$. 

\begin{figure}[ht]
	\includegraphics[width=8cm, height=7cm]{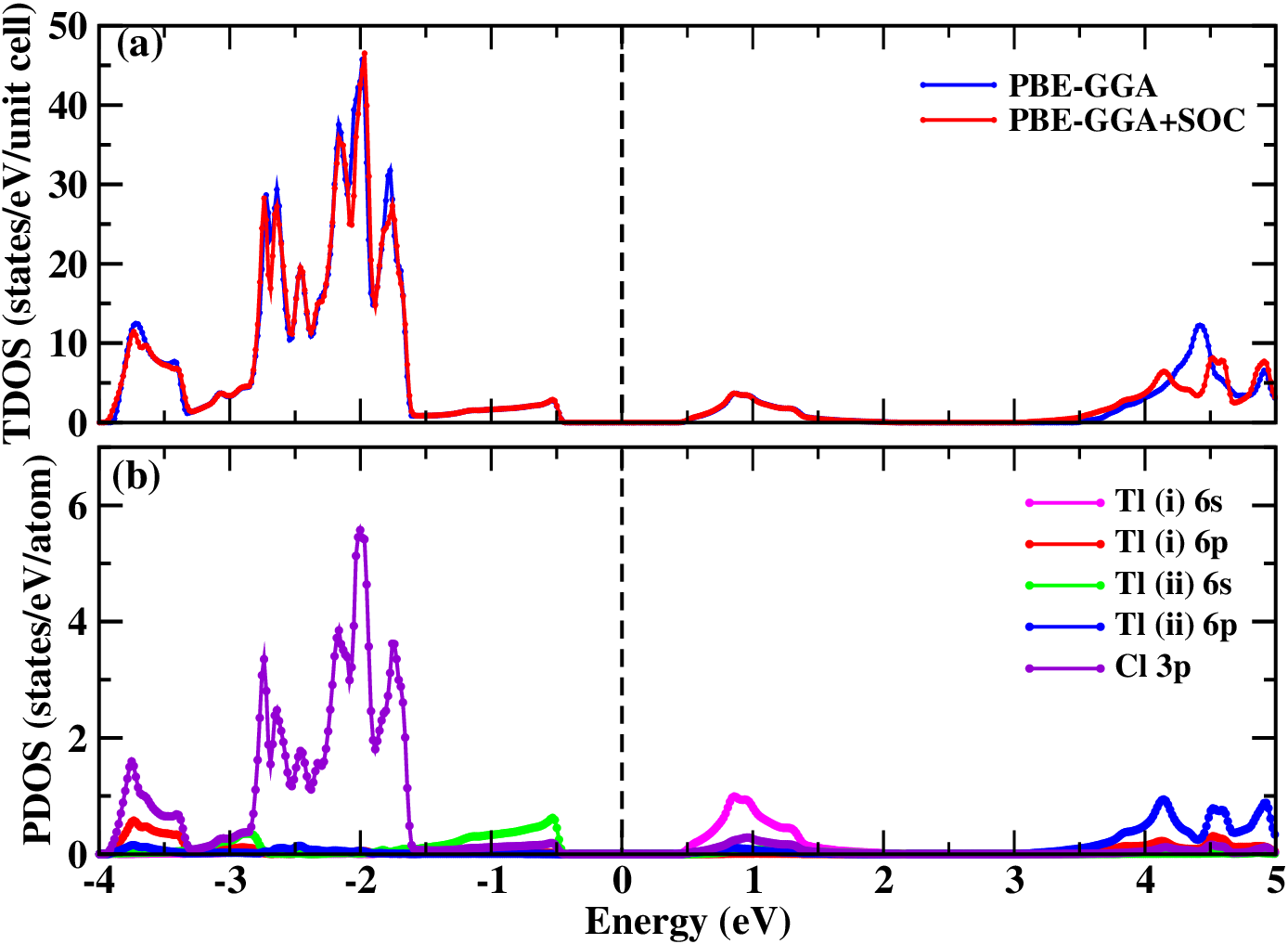} 
	\caption{Plots of (a) Total and (b) partial density of states (TDOS and PDOS) of RbTlCl$_3$.} 
\end{figure} 
To comprehensively investigate the contributions of constituent atoms, an analysis of the TDOS and PDOS has been conducted. The TDOS is depicted in Fig. 3(a), revealing the semiconducting nature of RbTlCl$_{3}$. It is clear from the plot of TDOS that both functionals are giving almost identical results. Within the VB region, a notable number of states are observed in the energy range between $\sim$-3 eV and -1.5 eV from two approaches. The maximum states in the VB are located around –2 eV, which can effectively contribute to the solar energy conversion process. After the band gap, the number of states in the CB is observed at $\sim$1 eV. A second set of peaks appears in the energy range of $\sim$3.5 to 5 eV. The TDOS from PBE-GGA+SOC and PBE-GGA has shown a quite identical trend. Moreover, the effect of SOC is evident only in the high-energy region (around 4.5 eV). It is noteworthy that the electronic dispersion and TDOS obtained in the presence and absence of SOC exhibit nearly identical characteristics except above $
\sim$3.5 eV. Therefore, we can chose any of the case, with or without SOC (here we select the SOC) for the calculations of the PDOS and both (PBE-GGA+SOC and PBE-GGA) for optical properties.

Furthermore, Fig. 3(b) depicts the PDOS. This is important in optical properties calculations. It is noticed in the figure that the major contribution in the VB is given by the 3p states of the Cl atom. The highest number of states is situated at the energy $\sim$-2 eV. The band gap in this material clearly arises from the contribution of Tl(i) 6s states in the CB side and Tl(ii) 6s states in the VB side. In the energy range from $\sim$1.9 eV to 3.5 eV, no states are available. Moving ahead in the CB, the 6p-states of Tl (i) and Tl (ii) have less contribution near E$_{F}$. Therefore, the transition process is likely dominated by the Cl 3p and Tl(ii) 6s states in the VB region and the Tl(i) 6s states in the CB. This information will be further utilized to analyze the contributions of specific electronic states at particular photon energies to transitions from the VB to the CB via photon absorption.

\subsection{\label{sec:optical}Optical properties}

In addition to electronic and phonon characteristics, it's crucial to take optical properties into account when assessing the effectiveness of solar absorber materials \cite{fox2002optical}. The optical properties of the materials encompass various phenomena that emerge when light interacts with them. To capture different levels of approximations, the imaginary part of the macroscopic dielectric function $\epsilon_{2}(\omega)$ is calculated using four different approximations. In the first approach, we have used the IPA, which only requires KS eigenvalues and eigenstates. In the second method, we apply the scissor correction over the IPA eigenvalues to obtain the optical spectra under the IQPA. In the third approach, the LFEs has been taken into account at RPA level. The fourth method, in contrast, uses the BSE to study effective e-h interactions \cite{onida2002electronic}. 

In Figs. 4(a) and 4(b), the variations of $\epsilon_{1}(\omega)$ and $\epsilon_{2}(\omega)$ are shown with the photon energy, respectively. These graphs have shown the results obtained from the IPA, IQPA, RPA, and BSE methods with and without SOC effects. The $\epsilon_{1}(\omega)$ describes how a material polarizes in response to an external electric field of light \cite{fox2002optical}. In Fig. 4(a), the static value of $\epsilon_{1}(\omega)$, $\epsilon_{1}(0)$, from the IPA is found to be $\sim$3.9. Whereas this value slightly decreases ($\sim$3.4) within IQPA method. This decreasing value, resulting from the increasing band gap introduced by the scissor correction, can be understood in terms of light-matter interaction: a larger band gap leads to lower polarization, which in turn results in a reduced dielectric constant \cite{ravindra2007energy}. Additionally, the inclusion of LFEs has given the value of $\sim$3.3, which is the lowest in all studied approaches. A similar reducing trend has also been observed in other indirect and direct band gap semiconductors such as Si, SiC, AlP, and GaAs, upon the introduction of LFEs \cite{gajdovs2006linear, hybertsen1987ab, baroni1986ab}. On adding the e-h attraction within BSE, the highest $\epsilon_{1}(0)$ is found to be $\sim$4. Here the low values of $\epsilon_{1}(0)$ as compared to Si \cite{hybertsen1987ab, baroni1986ab} indicate weak dielectric screening, which means the Coulombic attraction between electrons and holes is strong. As a result, the exciton binding energy is expected to be high \cite{PhysRevB.107.235119, solet2025many}.
\begin{figure}[ht]
	\includegraphics[width=8.3cm, height=8.9cm]{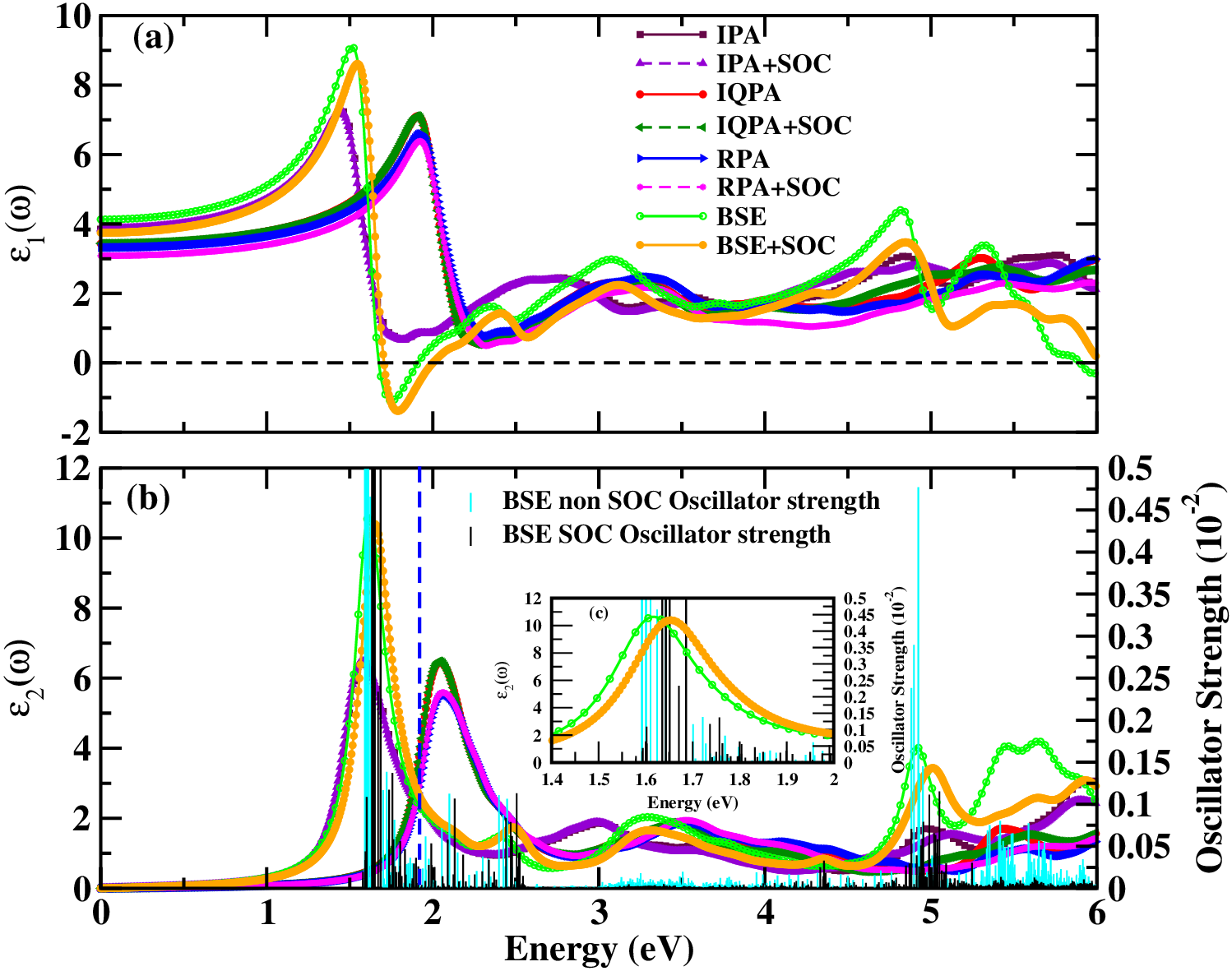}
\caption{Calculated (a) real $\epsilon_{1}(\omega)$ and (b) imaginary $\epsilon_{2}(\omega)$ parts of the dielectric function $\epsilon(\omega)$. The oscillator strengths of excitons calculated using the BSE method without and with SOC are shown by cyan and black vertical lines, respectively, in plot (b). The inset (c) presents an enlarged view of the energy range from 1.4 to 2.0 eV, providing a clearer visualization of the BSE-induced $\epsilon_{2}(\omega)$ spectra without and with SOC, together with their corresponding excitonic oscillator strengths.}
\end{figure}

A maximum peak value of $\epsilon_{1}(\omega)$ of $\sim$7.3 at $\sim$1.9 eV is obtained upon including the scissor correction over IPA (IQPA). The IQPA spectra has shown a rigid shift to higher energy compared to IPA. This value is slightly higher than that obtained using RPA, which also lies at $\sim$1.9 eV but with a somewhat lower value of $\sim$6.6. The highest value using the BSE is observed in the IR region of the solar spectrum, with a value of $\sim$9.1 at $\sim$1.52 eV. After these highest peaks, the BSE spectrum drops sharply and turns negative in the visible region, indicating metallic behavior from about $\sim$1.6 to 2 eV \cite{ashcroft1976solid}. This decrease arises due to excitonic effects, which are not captured in the other approximations. All four approaches exhibit oscillatory behavior throughout the studied energy range.

To understand the influence of SOC on the dielectric response, we now analyze its effect on the $\epsilon_{1}(\omega)$. The effect of SOC on $\epsilon_{1}(\omega)$ is negligible at the IPA and IQPA levels, as both approaches yield nearly identical results with and without SOC. In the case of RPA, a slight decrease in $\epsilon_{1}(0)$ is observed when SOC is included, from $\sim$3.4 to $\sim$3.2. A similar small reduction is seen in the BSE approach, where $\epsilon_{1}(0)$ has decreased from $\sim$4 to $\sim$3.8. This modest lowering of the static dielectric constant with SOC is also reflected in the spectral response, where the peak value of $\epsilon_{1}(\omega)$ from BSE in the IR region slightly decreases upon SOC inclusion. Overall, the effect of SOC remains minimal at the IPA and IQPA levels and becomes only marginally visible in the RPA and BSE spectra. In the context of solar cell applications, RbTlCl$_{3}$ has high value of the $\epsilon_{1}(\omega)$ in the IR region indicates strong polarizability of this material, which enhances its ability to store photon energy. 

The information of energy absorption in the material during the polarization process can be understood by $\epsilon_{2}(\omega)$. Figure 4(b) represents $\epsilon_{2}(\omega)$ and the oscillator strength of excitons. The optical spectra from IPA and IQPA have shown the same peak value of $\sim$6.5, as expected since IQPA is a rigidly shifted IPA spectrum, with the peak appearing at $\sim$1.6 and $\sim$2 eV in the near IR and visible region, respectively. When LFE is added, the broadening of $\epsilon_{2}(\omega)$ decreases, with a peak value of $\sim$5.6 around the scissor corrected direct gap. This reduction has also been observed in several other perovskites, such as BaTiO$_{3}$ and SrTiO$_{3}$ \cite{krasovskii1999local}. The BSE-calculated $\epsilon_{2}(\omega)$ spectrum reveals a noticeable redshift and sharper absorption features compared to IQPA and RPA, indicating significant excitonic effect. Notably, $\epsilon_{2}(\omega)$ spectra have reached a maximum value of $\sim$10.2 in the IR region at $\sim$1.6 eV. The $\epsilon_{2}(\omega)$ spectra have displayed oscillatory behavior, with distinct peaks and valleys arising from multiple interband transitions across the studied energy range. This material has shown maximum absorption in the IR region, which can be utilized in tandem solar cells \cite{rani2025hybrid}

The effect of SOC on the $\epsilon_{2}(\omega)$ spectra varies across the different levels of approximation. The influence of SOC is minimal in the case of IPA and IQPA. A minor deviation has been observed in the higher energy region (above $\sim$3 eV). The RPA spectra exhibit a comparable trend, with minimal SOC impact below $\sim$3 eV. In contrast, the BSE-calculated $\epsilon_{2}(\omega)$ spectrum shows the minor effect of SOC in the beginning; later on, this effect became significant in the high-energy region (above $\sim$4.5 eV). We have also observed this variation in the TDOS in the high-energy region. Therefore, this pattern has suggested a sensitivity to SOC when accounting for the excitonic effect. In this part, we shall discuss the impact of excitonic effects on $\epsilon_{2}(\omega)$. To further examine the intensity of these optical excitations, the oscillator strength has been calculated and analyzed. Figure 4(b) shows sharp peaks of the oscillator strength obtained from BSE, both in the presence and absence of SOC. Additionally, the inset in Fig.~4(c) presents an enlarged view of the 1.4-2.0 eV energy range, providing a clearer comparison of the BSE-induced $\epsilon_{2}(\omega)$ spectra with and without SOC, together with their corresponding excitonic oscillator strengths. This magnified view highlights the changes in the excitonic peak intensity and oscillator strengths induced by SOC. Excitons with high intensities are called bright excitons, and the excitons with the negligibly small intensities are called dark excitons. The maximum number of high-intensity excitons, or bright excitons, are observed below the scissor corrected direct band gap ($\sim$1.92 eV). The optical gap corresponds to the energy of the first exciton, which marks the onset of optical absorption in the BSE spectrum \cite{onida2002electronic}. According to this definition, the optical gap of the compound is found to be $\sim$1.6 eV in the non-SOC case. There are four excitons with the maximum probability of formation present at the energies of $\sim$1.59, $\sim$1.60, $\sim$1.61, and $\sim$1.64 eV with the oscillator strengths of $\sim$0.007, $\sim$0.009, $\sim$0.01, and $\sim$0.007, respectively. Their presence causes the highest value of $\epsilon_{2}(\omega)$ in this IR energy range, indicating that maximum absorption will take place in this energy region. One more bright exciton is observed around $\sim$4.9 eV with an intensity of $\sim$0.005, contributing to the second-highest value of the absorption spectrum in the UV energy region. The probability of exciton formation in the visible region is low compared to the IR and UV energy regions. Moreover, the dark excitons can be seen between the energy range of $\sim$2.5-4.2 eV. The binding energy of the $\lambda^{\text{th}}$ exciton in the excitonic state $A_{\nu ck}^{\lambda}$ can be defined as the difference between the exciton energy $E^{\lambda}$ and the scissor-corrected direct band gap. The binding energies of four excitons (discussed above) present at the high absorption peak obtained from the earlier definition are $\sim$365, $\sim$356, $\sim$347, and $\sim$315 meV, respectively. The high values of exciton binding energies indicate the presence of stable excitons at room temperature \cite{fox2002optical, haug2004quantum, peter2010fundamentals}. The localized (or delocalized) nature of these excitons will be analyzed in detail in the exciton property section. 

Overall, in the presence of SOC, the oscillator strength of excitons is reduced across the energy range, leading to lower-intensity excitonic peaks. However, the highest number of bright excitons still aligns with the strongest peak of $\epsilon_{2}(\omega)$ in the SOC-induced BSE spectrum. Since the SOC and non-SOC spectra have shown nearly the same peak values, we will focus on the SOC results in the analysis of other optical properties.

\begin{figure}[ht]
	\includegraphics[width=8cm, height=7cm]{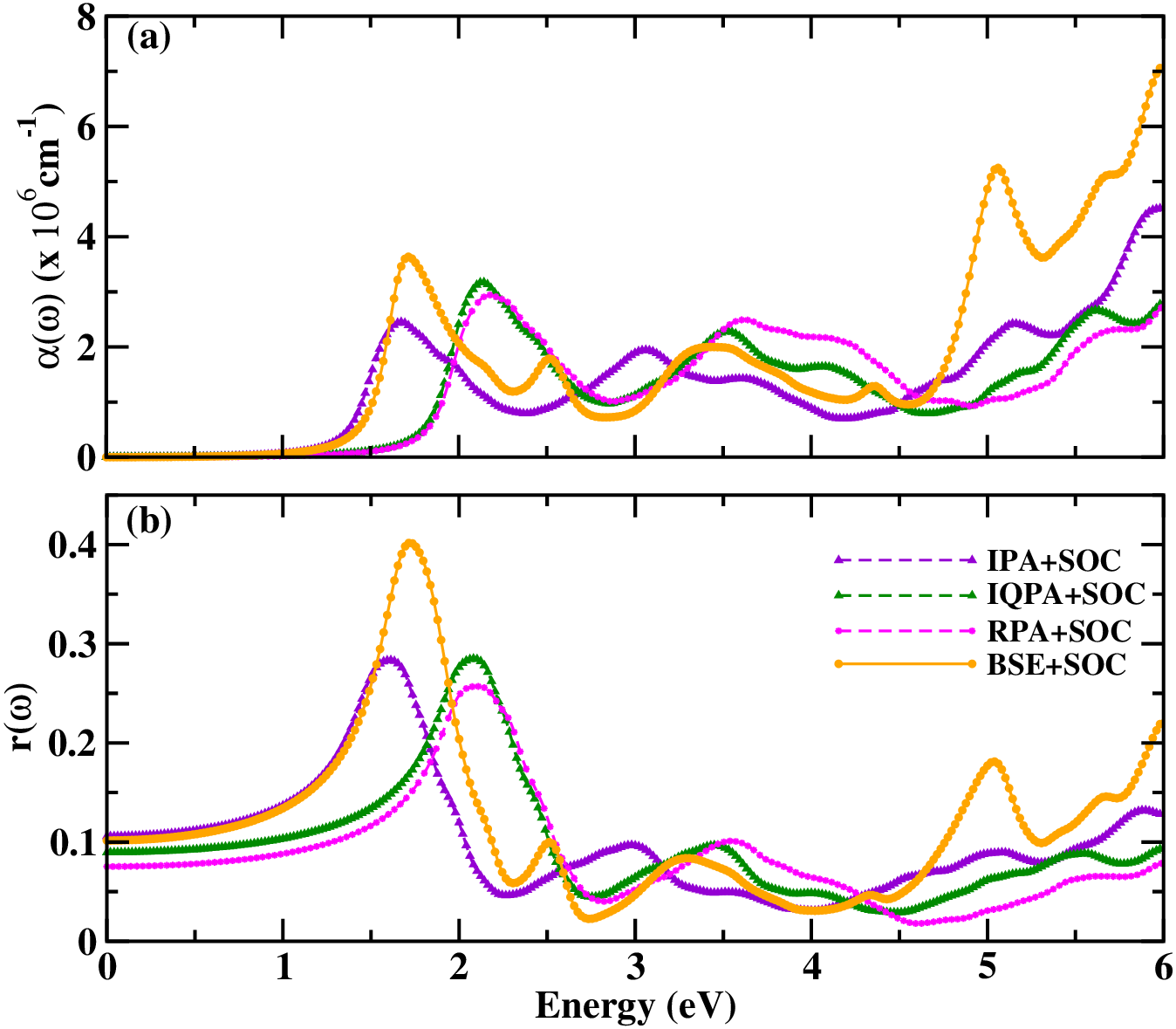} 
	\caption{The calculated (a) absorption coefficient \textbf{$\alpha(\omega)$} and (b) reflectivity r($\omega$) of RbTlCl$_3$ from IPA, IQPA, RPA, and BSE approaches with SOC.} 
\end{figure}
In this section of optical properties, we examine the variation of the absorption coefficient [$\alpha(\omega)$] and the reflectivity [$r(\omega)$], as functions of incident photon energy, evaluated using different levels of approximation. In particular, the $\alpha(\omega)$ will be used in the calculation of SLME. Figures 5(a) and 5(b) present the corresponding spectra of $\alpha(\omega)$ and $r(\omega)$, respectively. In Fig. 5(a), the absorption has been started before the edge of the minimum direct band gap from respective approximations (IPA, IQPA, RPA, and BSE) and reached a maximum value of $\sim$3.6$\times 10^{6}$ $cm^{-1}$ at $\sim$1.7 eV in the near visible region (from BSE). Similarly, $\alpha(\omega)$ obtained from IQPA (RPA) has the highest value of $\sim$3.2 (2.9)$\times 10^{6}$ $cm^{-1}$ at 2 eV in the visible light. After an intense energy point, peaks from IQPA and RPA show a slow decrease and then gradually increase with the photon energy. On the same graph, the BSE curve has multiple distinct peaks and valleys across the energy range, which shows oscillatory behavior in the spectra. The sharp rise near the optical gap is due to the presence of bound e-h pairs, which enhance absorption. The value reaches its peak in the IR region, and then it rises in the UV region. This trend reveals that the maximum absorption of photons will take place in these two regions. Han \textit{et al.} found that compounds from the same halide perovskite family, such as CsGeX$_3$ (X: Cl, Br, and I), exhibit $\alpha(\omega)$ values of an order of 10$^{5}$ cm$^{-1}$ \cite{thi2022electronic} by considering excitonic effects. The value of $\alpha(\omega)$ for RbTlCl$_3$ obtained in our work is comparable to that of CsGeX$_3$ \cite{thi2022electronic}. Figure 5(b) presents the plot of $r(\omega)$, which shows the fraction of light reflected by the material across different regions of the solar spectrum. Starting from their static values, the plots obtained from all four approximations rise and reach their respective maxima. After reaching their maxima, all peaks decrease sharply. The minimum value of $r(\omega)$ is observed to be $\sim$2.3 \% at energy $\sim$2.7 eV in the active region of the solar energy spectrum (from BSE). In contrast, the lowest $r(\omega)$ from IQPA (RPA) is observed at photon energy $\sim$4.5 eV with a value of $\sim$2.8 (1.8) \% in the UV region. Such very low value of $r(\omega)$ indicates that this material could be useful for the anti-reflecting coating. Beyond these minima, the $r(\omega)$ curves increase gradually as the photon energy extends further into the studied energy range. Overall, the plots from IQPA and RPA display nearly identical features. The compound RbTlCl$_3$ exhibits low r($\omega$) and high $\alpha(\omega)$ in the visible region, showing it would be useful for solar cell applications.

Furthermore, as the dielectric function, refractive index, absorption coefficient, and optical conductivity are closely related optical quantities. Therefore, the complex refractive index and optical conductivity are discussed here only at the BSE+SOC level, while their comprehensive comparison using the IPA+SOC, IQPA+SOC, RPA+SOC, and BSE+SOC approximations, along with the corresponding figures, is provided in the Supplementary Material (SM) \cite{SM}.

The refractive index of a material is a complex quantity that describes how light propagates through it. The real [n($\omega$)] and imaginary [k($\omega$)] parts of the complex refractive index describe the slowing down of light in the material and the absorption loss during its propagation, respectively. The corresponding results obtained using all four theoretical approximations are presented in Figs. S2(a) and S2(b) of the SM \cite{SM}. At the BSE+SOC level, a pronounced excitonic effect is observed, resulting in an enhanced static refractive index of $\sim$2.0 and a maximum value of n($\omega$) of about 3.1 in the IR region. The corresponding extinction coefficient, k($\omega$), exhibits a prominent peak of $\sim$2.0 near 1.68 eV, followed by a secondary peak in the UV region. The enhanced values of both n($\omega$) and k($\omega$) within the BSE+SOC framework indicate strong light--matter interaction arising from excitonic effects, highlighting the potential of RbTlCl$_3$ for photovoltaic applications.

Similarly, the optical conductivity provides further insight into the optical response of RbTlCl$_3$. It is a well-established fact that electrons get excited from the VB to the CB by absorbing photons. This transition between the two bands is commonly referred to as optical conduction, driven by the absorption of light \cite{ashcroft1976solid}. This can be explained by the $\omega$ dependent complex optical conductivity, which includes the real [$\sigma_{1}(\omega)$] and imaginary [$\sigma_{2}(\omega)$] parts. The $\sigma_{1}(\omega)$ and $\sigma_{2}(\omega)$ are associated with the $\epsilon_{2}(\omega)$ and $\epsilon_{1}(\omega)$, which are related to the absorption of material and the amount of polarization \cite{dressel2002electrodynamics}. The optical conductivity spectra calculated within the IPA+SOC, IQPA+SOC, RPA+SOC, and BSE+SOC approximations are presented in Figs. S3(a) and S3(b) of the SM \cite{SM}. The BSE+SOC results reveal a pronounced excitonic enhancement in both the real [$\sigma_{1}(\omega)$] and imaginary [$\sigma_{2}(\omega)$] parts of the optical conductivity. The $\sigma_{1}(\omega)$ spectrum exhibits strong absorption below the direct scissor operated band gap, with the first and second prominent peaks of approximately $5.0\times10^{15}$ s$^{-1}$ and $5.1\times10^{15}$ s$^{-1}$ located at around 1.6 and 4.9 eV, respectively. These peaks coincide with the energies of the bright excitons, indicating strong excitonic contributions to the optical response. Likewise, $\sigma_{2}(\omega)$ shows a pronounced peak of approximately $3.5\times10^{15}$ s$^{-1}$ near 1.5 eV, demonstrating enhanced polarization arising from excitonic effects. The positive value of $\sigma_{2}(\omega)$ near the optical onset indicates capacitive behavior associated with energy storage through polarization. Overall, the enhanced optical conductivity within the BSE+SOC framework highlights the strong light-matter interaction in RbTlCl$_3$, reinforcing its potential for photovoltaic application.

\subsection{\label{sec:exciton}Excitonic properties} 
The nature of excitons formed due to the optical excitation can be comprehensively assessed both in reciprocal and real space. Solving the BSE enables the excitation spectrum and a detailed understanding of the exciton's spatial properties.
\begin{figure}[ht]
	\includegraphics[width=8.5cm, height=7cm]{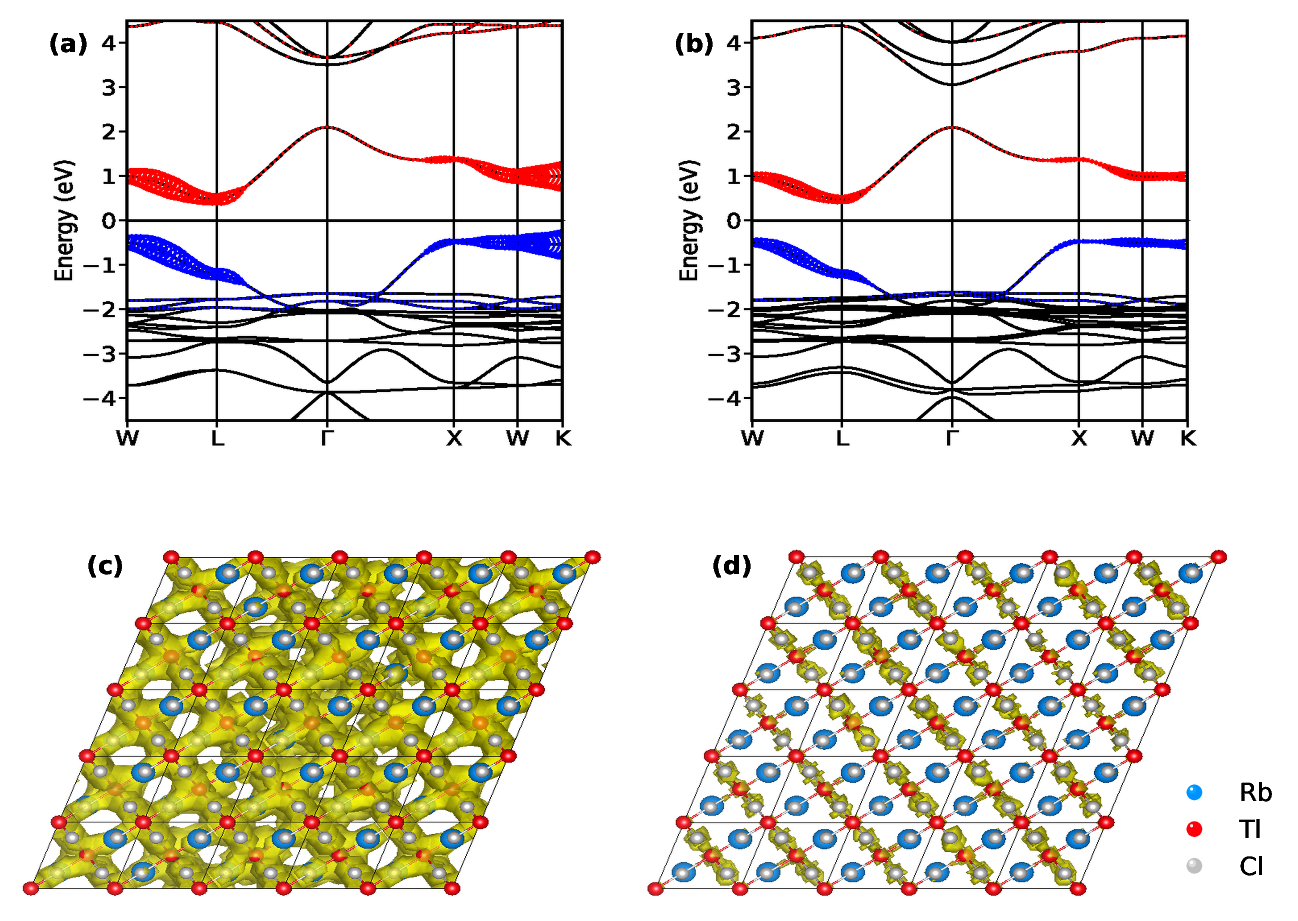} 
	\caption{The exciton distribution in reciprocal space without (a) and with (b) SOC is represented by blue and red circles, respectively, superimposed on the band structure. The radius of each circle is proportional to the squared amplitude of the $\lambda^{\text{th}}$ excitonic wave function. The corresponding real-space distribution of the high-intensity excitonic state without (c) and with SOC (d) is associated with the highest peak in the $\epsilon_{2}(\omega)$ spectrum. These distributions are calculated using a 5$\times$5$\times$5 supercell and visualized along the $\mathbf{c}$ axis. The isosurface level is set to 6.2 $\times 10^{-7}$, as visualized in VESTA. The hole is fixed near the Tl atom at the fractional coordinates (0.52, 0.52, 0.52).} 
\end{figure}
Here, we focus on the exciton with the highest oscillator strength in the studied energy range, as it dominates in the absorption spectrum by contributing significantly to the highest peak in $\epsilon_{2}(\omega)$ and thus represents the optically active transition in the material. In Figs. 6(a) and 6(b), the exciton weights are overlaid on the rigidly shifted DFT band structure. The larger blue and red circle radii indicate a stronger contribution of the respective valence- and conduction-band states at a given k-point to the corresponding BSE excitonic eigenstate. In these figures, the hole and electron distributions of the $\lambda^{\text{th}}$ bound e-h pair in momentum space are described by Eqs. 6(a) and 6(b) for the valence and conduction bands, respectively. In the absence and presence of SOC, the exciton distribution in momentum space is concentrated along the $W$–$L$ and $X$–$W$–$K$ paths in the Brillouin zone, indicating its localized character in momentum space. Conversely, in real space, their nature is expected to be delocalized. The figure also indicates that the intense peak in $\epsilon_{2}(\omega)$ originates primarily from the optical gap region, where the probability of forming bright excitons is highest. The overall distribution of the excitonic weights in reciprocal space remains qualitatively similar upon inclusion of SOC; however, noticeable changes occur in the relative magnitudes of the weights associated with the valence and conduction bands.

Furthermore, the localization (or delocalization) of excitons in real space is intrinsically linked to the distribution of their corresponding wave functions in reciprocal space. To better understand this behavior, we examine a real-space representation of the excitonic wave function associated with a prominent peak in the absorption spectrum. As discussed earlier, the wave function of an exciton depends on both electron and hole positions, forming a six-dimensional object. To visualize it, we locate the hole near the Tl site at (0.52, 0.52, 0.52) and plot the exciton wave function with respect to the coordinates of the electron. To illustrate the spatial profile of the exciton in real space, we employ the VESTA software \cite{momma2008vesta} and choose an isosurface value of 6.2 $\times 10^{-7}$. Figures 6(c) and 6(d) show the three-dimensional view of the interacting e-h wave function in the 5$\times$5$\times$5 supercell viewed along the $\mathbf{c}$ axis without and with SOC, respectively. These figures illustrate the spatial distribution of the excited electron corresponding to the first optically allowed excitonic state using an isosurface representation. In Fig. 6(c), a significant distribution of the wave function can be seen on the Cl and Tl sites in the 3$\times$3$\times$3 part of the supercell. The spread of the exciton wave function across the unit cells of the supercell reveals the delocalized character of the exciton. Similarly, in Fig. 6(d), it can be observed that the localization of this high-intensity exciton is at the Cl and Tl sites, whereas it vanished at the Rb site. It is clear from this plot that the exciton wave function is not confined within one unit cell; rather, it is distributed all over the space. The main difference between the non-SOC and SOC cases is the change in the spatial distribution and relative amplitude of the excitonic wave function across the unit cells. This analysis indicates that the excitons in the material are delocalized with the binding energy ranging from $\sim$0.315 to 0.365 eV. The reason for large exciton binding energies in this material arises predominantly from reduced dielectric screening, resulting the bound exciton is spatially extended. 

An important consideration for PV applications is that excitonic absorption does not immediately generate free charge carriers. Upon photon absorption, the resulting bound e-h pair has to first dissociate into free carriers before contributing to the current. However, during this process, there is the possibility of recombination, whereby the electron returns to the hole before being collected, resulting in losses of pairs. Here, the spatial extent of the exciton could play a crucial role in reducing these pair losses. In RbTlCl$_{3}$, the exciton wave function is delocalized. Such spatial delocalization can facilitate carrier separation by allowing the resulting free carriers to be predominantly present in the neutral region of a pn junction solar cell, where the probability of recombination is considerably lower \cite{nelson2003physics}. Interestingly, similar delocalized excitons have also been reported in perovskite Cs$_{3}$Bi$_{2}$Br$_{6}$I$_{3}$ \cite{kaur2025crossover}. Therefore, the delocalized nature of the excitons in real space is a favorable characteristic that may benefit future PV and optoelectronic applications of RbTlCl$_{3}$.  

\subsection{\label{sec:SLME}Spectroscopic limited maximum efficiency (SLME)}   
Evaluating material performance is key to designing efficient solar cells. While PCE is a common metric, this study has used SLME for the assessment of PV potential. In comparison with the SQ model \cite{shockley1961detailed}, SLME considers the material's true absorption, the type of band gap (direct and indirect), and the thickness of the absorber layer. The PCE from SLME has been evaluated under the standard AM1.5 solar spectrum at a temperature of 300 K \cite{yu2012identification, nrel-am15}.

\begin{figure}[ht]
	\includegraphics[width=8cm, height=7cm]{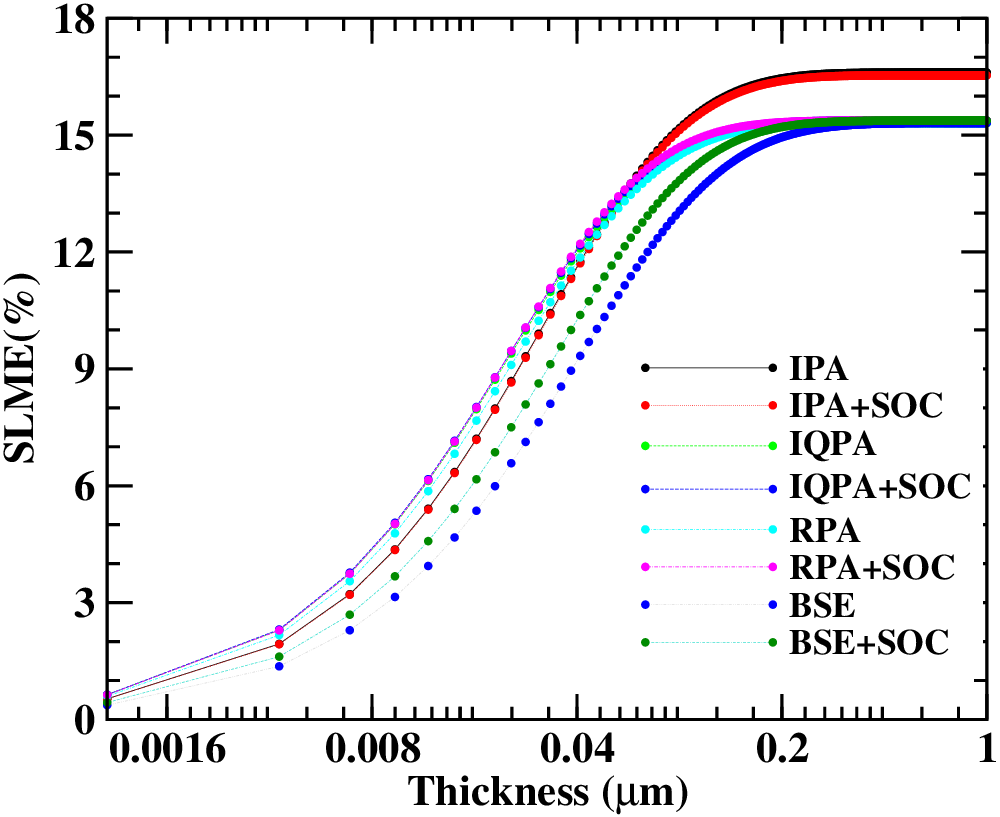} 
	\caption{The calculated SLME with four different methods for RbTlCl$_3$ against the thickness.} 
\end{figure} 
In Fig.~7 we have presented the variation of SLME of single junction RbTlCl$_{3}$ with the absorber layer thickness from 0 to 1 $\mu$m under four approximations in the presence and absence of SOC. The effect of SOC has marginally increased SLME, especially at lower thicknesses. This increase has been attributed to changes caused by SOC in the band structure and optical transitions. For all four approximations (IPA, IQPA, RPA, and BSE), the SLME has increased rapidly from 0 \% to around 15.4-16.6 \% as the thickness has increased up to $\sim$0.5 $\mu$m. Beyond this thickness range, SLME has been observed to be saturated, indicating that the absorber layer has become thick enough to capture photons. Here IPA yields the highest SLME ($\sim$16.6 \%) at a thickness of $\sim$0.4 $\mu$m, while IQPA, RPA, and BSE give slightly lower values (15.4 \%) at $\sim$0.5 $\mu$m. This difference stems from $f_{r}$ being higher than the other three methods, which is expected. The same values of SLME was expected, because all three approaches used the same minimum direct and fundamental gaps. In conclusion, all approximations have predicted similar saturation efficiencies, but BSE has provided a more accurate estimate, likely due to its more realistic treatment of excitonic effects. Furthermore, the SLME of 15.4 \% is higher than that of other indirect band gap semiconductors, such as Cs$_{2}$In$_{2}$X$_{6}$ (X: F, Br) with SLME values of 0.1 \% and 11.5 \%, respectively~\cite{palummo2020optical}, and Cs$_{2}$AgBiX$_{6}$ (X: Cl, Br, I) with SLME values ranging from 3.9 \% to 12.37 \%~\cite{palummo2020halide}. In addition, a similar class of indirect perovskite material, TlSiF$_{3}$, exhibits an SLME almost twice that of the studied compound~\cite{yadav2022study}. A few other perovskites with high SLME include Cs$_{2}$AgCoX$_{6}$ (X: Cl, Br) \cite{nabi2024lead} and Cs$_{2}$AgCrX$_{6}$ (X: Cl, Br) \cite{nabi2024lead}, both with efficiency ranges of $\sim$23-31.7\%. It is worth noting that the SLME of RbTlCl$_{3}$ is calculated using the $\alpha(\omega)$ obtained from the BSE approach, providing a more accurate description of the optical absorption, whereas the reported SLME values for the above compounds were obtained using $\alpha(\omega)$ from the IPA. These observations suggest that RbTlCl$_{3}$ might be a promising candidate for potential thin-film PV applications and provide motivation for future experimental investigations.

In semiconducting PV materials, a variety of recombination processes can occur~\cite{nelson2003physics}, such as radiative and non-radiative recombinations. But in the SLME formalism, only radiative recombination is explicitly considered. This recombination occurs via a single-step direct process in which carriers recombine after band-to-band transitions. In other way, the other types of recombinations, such as non-radiative, can also be particularly valuable when making real solar devices. This process is multi-step, where multiple types of defect (as shown in Table~\ref{tab:RbTlCl3_DFE}) can help in completing the process, meaning that it will be completed through an external agent. Based on this, one can say that the current due to latter recombination is comparatively lower than the former one-step process inside solar cells. For indirect band gap materials, the recombinations other than radiative are considered through $f_r$. It means that in our studied RbTlCl$_3$ material, we can say that other recombinations are already included within SLME calculations. However, in many previous studies for indirect gap semiconductors, it has been noticed that the change in efficiency by including non-radiative recombination via defect calculations is negligible. For example, in Mg$_2$Si, the maximum SLME of 1.3 \% is only changed to 1.2 \% due to non-radiative recombination~\cite{solet2024mg}. Moreover, defect studies for selenium semiconductor revealed no signs of non-radiative recombination in the efficiency computation~\cite{moustafa2024selenium}. Thus, one can expect the efficiency of more than 15 \% in RbTlCl$_3$ if we will also consider the non-radiative carrier recombination by using defect properties of this paper.

The electronic, optical, excitonic, and SLME properties of RbTlCl$_3$ have been discussed in the preceding sections. However, since the compound has not yet been experimentally synthesized, it is important to examine its stability before discussing its experimental realization. Therefore, phonon calculations have been performed to verify its dynamical stability, followed by an analysis of the thermodynamic phase stability and native defects, which provides insight into the feasibility of synthesizing RbTlCl$_3$.

\subsection{Dynamical Stability, Phase Stability, and Native Defects}
\label{sec:stability}
The assessment of dynamical stability is essential for evaluating the potential experimental realization of RbTlCl$_3$. To analyze its dynamical stability, phonon calculation has been performed using the optimized lattice parameter and fully relaxed atomic positions obtained from the structural optimization \cite{SM}. Dynamical stability is examined through the phonon dispersion relations calculated along the high-symmetry directions of the first Brillouin zone. These vibrational characteristics not only confirm the structural stability of the compound but also provide the foundation for evaluating its thermodynamic properties.
\begin{figure}[ht]
	\includegraphics[width=8cm, height=6.7cm]{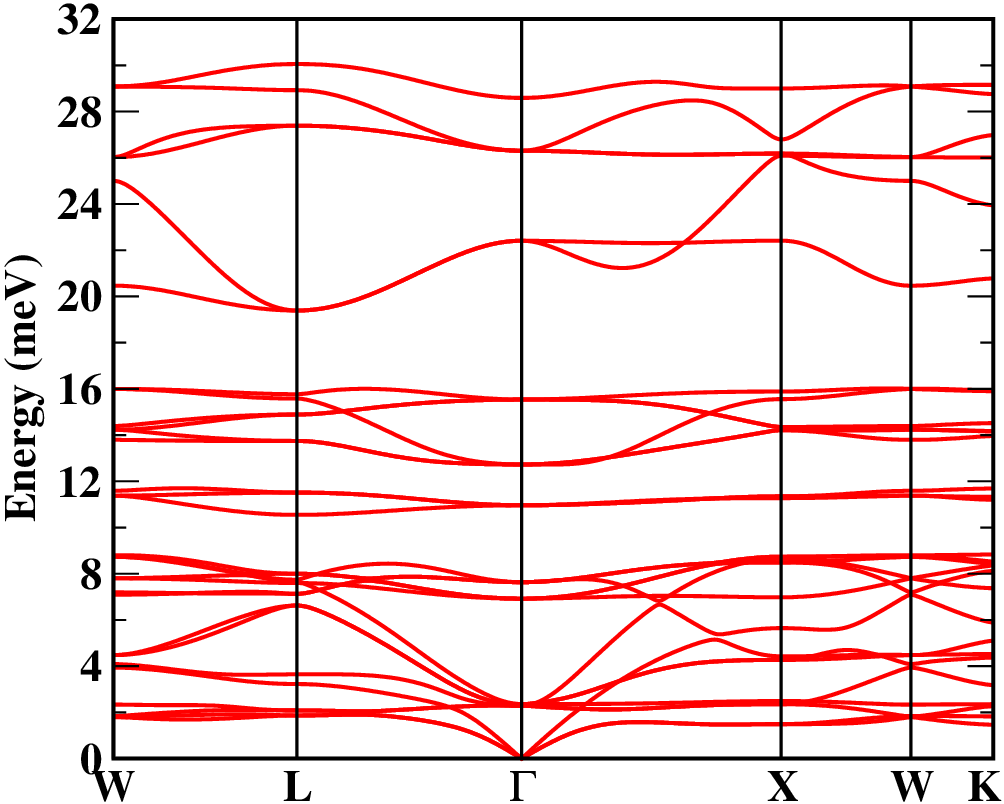} 
	\caption{Phonon dispersion of RbTlCl$_{3}$ in the high symmetric direction.} 
\end{figure} 
Figure 8 show the phonon dispersion of RbTlCl$_{3}$. The phonon dispersion is calculated along the high-symmetric $\textbf{k}$-path direction W-L-$\Gamma$-X-W-K in the first Brillouin zone. In Fig. 8, there are a total of thirty normal modes. Out of these thirty branches, first three branches in the energy range $\sim$0-2.3 meV are known as the acoustic phonon branches, which describe vibrations with energies that vanish linearly with $\textbf{k}$ in the long-wavelength limit \cite{ashcroft1976solid}. All these three branches are triply degenerate at the $\Gamma$ point. Further, the acoustic branch having the steepest slope is called the longitudinal acoustic (LA) branch, and the other two remaining branches having the smaller slopes are called transverse acoustic (TA) branches. The TA branches are doubly degenerate in the high-symmetric directions L-$\Gamma$ and $\Gamma$-X. There are twenty-seven optical branches out of thirty normal modes, situated in the energy range of $\sim$2.4-31 meV; their energies do not vanish in the long-wavelength limit \cite{ashcroft1976solid}. The phonon dispersion exhibits only positive phonon energies, indicating that the studied structure of RbTlCl$_{3}$ is dynamically stable within the harmonic approximation.

After establishing the dynamical stability of RbTlCl$_3$, its thermodynamic stability is examined through chemical potential analysis. As shown in Eq. (12), the defect formation energy depends on the chemical potentials of the constituent elements. These chemical potentials not only play a crucial role in determining the defect formation energy but also serve as a link between theoretical simulations and experimental growth conditions. It is well known that the chemical potential of an element is defined as the change in Gibbs free energy with respect to the amount of that element \cite{buckeridge2014automated}. A higher chemical potential of element \textit{i} in the synthesis environment corresponds to a higher molar concentration of that element in the material, i.e., a more \textit{i}-rich condition. At the computational level, the chemical potentials can be determined by analyzing the thermodynamic stability region for the synthesis of RbTlCl$_3$, rather than through the competing phases and compounds observed experimentally. Therefore, thermodynamic stability is essential to ensure the formation of RbTlCl$_3$ without decomposition into competing phases.

To synthesize RbTlCl$_3$ and ensure its stability under equilibrium conditions, the chemical potentials of the constituent elements in the growth environment, $\mu_{\mathrm{Rb}}, \mu_{\mathrm{Tl}}$, and $\mu_{\mathrm{Cl}}$, must satisfy the following conditions \cite{buckeridge2014automated}:
\begin{equation}
	\mu_{\mathrm{Rb}} + \mu_{\mathrm{Tl}} + 3\mu_{\mathrm{Cl}}
	=
	\Delta H_f,
\end{equation}
where $\Delta H_f$ is the calculated formation energy of RbTlCl$_3$, i.e., the energy change of the reaction Rb + Tl + 3Cl $\rightarrow$ RbTlCl$_3$. To prevent the formation of pure elemental phases during synthesis, the chemical potentials must satisfy the conditions $\mu_i$ $\le$ 0 (where $i$ = Rb, Tl, and Cl). On the other hand, to prevent the formation of secondary phases such as RbCl, TlCl, and Rb$_3$TlCl$_6$, as well as any phases that deviate from the stoichiometry (1:1:3) of RbTlCl$_3$, the following conditions must be satisfied \cite{buckeridge2014automated}:
\begin{equation*}
	\begin{aligned}
		\mu_{\mathrm{Rb}} + \mu_{\mathrm{Cl}}
		&< \Delta H_f(\mathrm{RbCl})  \\
		\mu_{\mathrm{Tl}} + \mu_{\mathrm{Cl}}
		&< \Delta H_f(\mathrm{TlCl}) \\
		\mu_{\mathrm{Tl}} + 3\mu_{\mathrm{Cl}}
		&< \Delta H_f(\mathrm{TlCl_3})  \\
		3\mu_{\mathrm{Rb}} + \mu_{\mathrm{Tl}} + 6\mu_{\mathrm{Cl}}
		&< \Delta H_f(\mathrm{Rb_3TlCl_6}).
	\end{aligned}
\end{equation*}
The pure phase of RbTlCl$_{3}$ can be synthesized and remain thermodynamically stable only when all these conditions are satisfied. These constraints limit the allowed values of $\mu_{\mathrm{Rb}}, \mu_{\mathrm{Tl}}$, and $\mu_{\mathrm{Cl}}$. By solving these inequalities, the range of chemical potentials within which the investigated perovskite is stable can be determined. In this work, $\mu_{\mathrm{Rb}}$ and $\mu_{\mathrm{Tl}}$ are treated as independent variables, while $\mu_{\mathrm{Cl}}$ is considered a dependent variable. The intersection of these constraints provides the thermodynamically stable region in chemical potential space, which is represented as a two-dimensional (2D) stability diagram with two independent chemical potentials \cite{buckeridge2014automated}.The obtained $\mu_{i}$, when combined with the chemical potential of the corresponding species in its stable state, can be used to calculate the defect formation energies of various intrinsic point defects \cite{buckeridge2014automated}.
\begin{figure}[ht]
	\label{fig:TDe}
	\includegraphics[width=8cm, height=6cm]{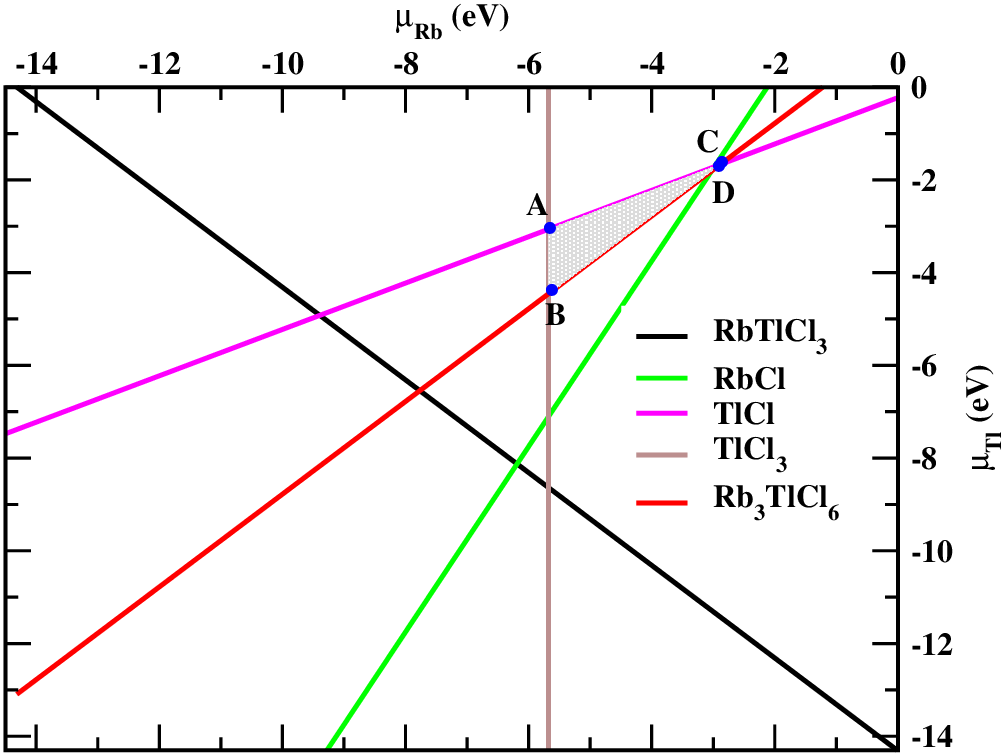} 
	\caption{The stability region (shaded) of RbTlCl$_{3}$ within the two-dimensional chemical potential space defined by $\mu_{Rb}$ and $\mu_{Tl}$, with coloured boundaries corresponding to phase-stability constraints from competing compounds and the target phase.} 
\end{figure} 

\begin{table}[h]
	\caption{Enthalpies of formation ($\Delta H_f$) of RbTlCl$_3$ and its relevant competing phases, obtained from DFT calculations.}
	\label{tab:RbTlCl3_formation}
	\centering
	\begin{tabular}{l c l c}
		\hline\hline
		Systems & $\Delta H_f$ (eV) & Systems & $\Delta H_f$ (eV) \\
		\hline
		RbTlCl$_3$      & $-14.31$ &  Rb$_3$TlCl$_6$ & $-29.85$\\
		TlCl            & $-4.92$  & TlCl$_3$    & $-8.63$ \\
		RbCl            & $-6.19$  & \\
		\hline\hline
	\end{tabular}
\end{table}

Figure 9 presents the thermodynamic stability region of RbTlCl$_{3}$ in the 2D chemical-potential $\mu_{Rb}$ and $\mu_{Tl}$ space, plotted using the enthalpies of formation listed in Table~\ref{tab:RbTlCl3_formation}. The shaded polygon represents the range of chemical potentials for which RbTlCl$_3$ is thermodynamically stable against decomposition into elemental and competing compound phases. Each boundary of the stability region corresponds to an equality condition between the formation enthalpy of RbTlCl$_3$ and that of a competing phase, thereby enforcing phase-stability constraints. The black line denotes the stability limit of RbTlCl$_3$, while the colored lines represent constraints imposed by competing phases, including RbCl (green), TlCl (magenta), TlCl$_3$ (brown), and Rb$_3$TlCl$_6$ (red). In Fig.~9, TlCl$_3$ forms one of the boundaries of the allowed chemical-potential region. The intersections of these constraints define the vertices of the stability region and are labeled A--D in Fig.~9. The corresponding chemical potentials and competing phases at each intersection are listed in Table~\ref{tab:RbTlCl3_intersections}.
\begin{table}[h]
	\caption{Intersection points bounding the stability region of RbTlCl$_{3}$ in $\mu_{\mathrm{Rb}}$ and $\mu_{\mathrm{Tl}}$ space, together with the corresponding $\mu_{\mathrm{Cl}}$ values and competing phases. All values are in eV.}
	\label{tab:RbTlCl3_intersections}
	\centering
	\begin{tabular}{c c c c l}
		\hline\hline
		Points &  $\mu_{\mathrm{Rb}}$ & $\mu_{\mathrm{Tl}}$ & $\mu_{\mathrm{Cl}}$ & Competing phases \\
		\hline
		A & $-5.68$ & $-3.06$ & $-1.86$  & TlCl, TlCl$_3$ \\
		B & $-5.68$ & $-4.46$ & $-1.39$  & TlCl$_{3}$, Rb$_3$TlCl$_6$ \\
		C & $-2.98$ & $-1.72$ & $-3.20$ & RbCl, TlCl \\
		D & $-3.03$ & $-1.81$ & $-3.16$ & RbCl, Rb$_3$TlCl$_6$ \\
		\hline\hline
	\end{tabular}
\end{table}

Point A marks the intersection between the TlCl$_3$ and TlCl stability boundaries, while Point B corresponds to the coexistence of TlCl$_3$ and Rb$_3$TlCl$_6$. Points C and D arise from the intersections of the RbCl-TlCl and RbCl-Rb$_3$TlCl$_6$ phase boundaries, respectively. Since the elemental Cl-rich condition ($\mu_{\mathrm{Cl}} = 0$ eV) lies outside the thermodynamic stability region due to competing phases, the Cl-rich limit is taken as the maximum allowable chlorine chemical potential, $\mu_{\mathrm{Cl}} = -1.39$ eV, corresponding to Point B. Together, these points delineate the full stability window of RbTlCl$_3$ in chemical-potential space.

Overall, the figure shows that RbTlCl$_3$ is thermodynamically stable only within a narrow (grey) region constrained by multiple competing phases, showing the importance of carefully controlled growth conditions to avoid the formation of other competing compounds.

To understand the defect behavior in RbTlCl$_3$, twelve intrinsic \textbf{neutral} point defects have been considered, including three interstitials ($I_{Rb}$, $I_{Tl}$, and $I_{Cl}$), six antisite defects ($Rb_{Tl}$, $Tl_{Rb}$, $Rb_{Cl}$, $Tl_{Cl}$, $Cl_{Rb}$, and $Cl_{Tl}$), and three vacancies ($V_{Rb}$, $V_{Tl}$, and $V_{Cl}$). For a comprehensive assessment of the defects in RbTlCl$_3$ and their role in its phase stability, charged defects should also be considered in addition to neutral point defects. This requires calculations for multiple possible charge states of each defect, determination of the Fermi-level dependence of the defect formation energies, and appropriate treatment of finite-size effects and charge corrections associated with charged supercells. Such an investigation is beyond the scope of the present study. Therefore, the present analysis is restricted to neutral point defects. The calculated defect formation energies under Cl-rich and Cl-poor growth conditions are summarized in Table~\ref{tab:RbTlCl3_DFE}. Under Cl-rich growth conditions, $V_{Tl}$ and $I_{Cl}$ possess the lowest positive DFEs with values of $\sim$0.07 and $\sim$0.46 eV/atom, respectively. This shows that these neutral point defects are energetically more favorable in this growth environment. Additionally, three defects show negative DFEs, namely the antisite defects ($Tl_{Rb}$ and $Cl_{Rb}$) and the cation vacancy ($V_{Rb}$). Usually, negative formation energies are considered unphysical. However, here it is important to note that the DFEs have been evaluated within the dilute-defect approximation, where defect-defect interactions are neglected. Due to the low dielectric constant of RbTlCl$_3$ (see Fig.~4(a)), this interaction term could be important for a more accurate evaluation of DFEs \cite{meggiolaro2018first}. Therefore, one should keep in mind this aspect when interpreting these results in future studies. Negative DFEs have also been reported in other halide perovskites \cite{xue2022intrinsic, liang2021first, liang2021electrode}. Importantly, defect formation energies depend strongly on the chemical potentials during synthesis. One can notice that, by changing the growth conditions from Cl-rich to Cl-poor environments, the negative formation energies of $Cl_{Rb}$ and $V_{Rb}$ become positive, which limits excessive defect formation \cite{xue2022intrinsic}. However, $Tl_{Rb}$ remains slightly negative under the Cl-poor condition.

\begin{table}[h]
	\caption{The obtained defect formation energies (DFEs) in eV/atom for different
		types of neutral point defects in RbTlCl$_3$ under Cl-rich and Cl-poor growth conditions.}
	\label{tab:RbTlCl3_DFE}
	\centering
	
	\begin{tabular*}{\columnwidth}{@{\extracolsep{\fill}}lcc}
		\hline\hline
		\shortstack{Defects in\\RbTlCl$_3$} & \multicolumn{2}{c}{DFE} \\
		\cline{2-3}
		& Cl-rich & Cl-poor \\
		\hline\hline
		
		$I_{\mathrm{Rb}}$ & 5.69 & 2.99 \\
		$I_{\mathrm{Tl}}$ & 5.61 & 2.87 \\
		$I_{\mathrm{Cl}}$ & 0.46 & 2.28 \\
		$\mathrm{Rb}_{\mathrm{Tl}}$ & 3.14 & 3.19 \\
		$\mathrm{Tl}_{\mathrm{Rb}}$ & -0.10 & -0.15 \\
		$\mathrm{Rb}_{\mathrm{Cl}}$ & 11.31 & 6.79 \\
		$\mathrm{Tl}_{\mathrm{Cl}}$ & 10.29 & 5.73 \\
		$\mathrm{Cl}_{\mathrm{Rb}}$ & -1.53 & 2.98 \\
		$\mathrm{Cl}_{\mathrm{Tl}}$ & 1.52 & 6.08 \\
		$V_{\mathrm{Rb}}$ & -2.00 & 0.69 \\
		$V_{\mathrm{Tl}}$ & 0.07 & 2.81 \\
		$V_{\mathrm{Cl}}$ & 2.22 & 0.41 \\
		
		\hline\hline
	\end{tabular*}
\end{table}

In contrast, under Cl-poor growth conditions, the $V_{Cl}$ and $V_{Rb}$ exhibit the lowest positive formation energies of $\sim$0.41 and $\sim$0.69 eV/atom, respectively, showing that these defects are more favorable in the Cl-poor environment. In comparison, other defects such as interstitials ($I_{Rb}$, $I_{Tl}$, and $I_{Cl}$) and antisites like $Rb_{Cl}$, $Cl_{Tl}$ and $Tl_{Cl}$ possess considerably higher formation energies and are therefore less likely to form. Overall, this analysis has shown that $V_{Cl}$ and $V_{Rb}$ are the most stable defects with positive formation energies under Cl-poor growth conditions. Therefore, controlling the chlorine chemical potential during synthesis may provide an effective way to limit defect populations and influence the defect landscape of RbTlCl$_3$.

The present theoretical results provide motivation for future experimental studies on RbTlCl$_3$. If successfully synthesized, its predicted electronic, optical, excitonic properties, and SLME can be experimentally verified, and its potential for PV applications can also be explored.
\section{Conclusions}
In this work, we have presented a detailed analysis of the electronic and optical properties of RbTlCl$_{3}$ using first-principles calculations, from ground-state DFT to excited-state methods, with and without SOC. The fundamental (direct) gap of this material are found to be $\sim$0.94 (1.49) and $\sim$0.95 (1.50) eV from PBE-GGA+SOC and PBE-GGA, respectively. Further, the optical spectrum has been calculated using four different approximations (IPA, IQPA, RPA, and BSE) to comprehensively study the effects at various levels of correction. The highest peaks of $\epsilon_{1}(\omega)$ and $\epsilon_{2}(\omega)$ from the BSE spectrum occur below the scissor corrected direct band gap, with values of $\sim$9.1 and $\sim$10.2 at photon energies of $\sim$1.5 eV and $\sim$1.6 eV, respectively. The corresponding maximum values of $\epsilon_{1}(\omega)$ ($\epsilon_{2}(\omega)$) from IQPA and RPA are $\sim$7.3 (7) and $\sim$6.6 (5.6), respectively, around the optical gap. Four bright excitons corresponding to the highest peak of $\epsilon_{2}(\omega)$ have been obtained from the BSE calculation, with binding energies of $\sim$365, $\sim$356, $\sim$347, and 315 meV in the IR region. Furthermore, the maximum value of $\alpha(\omega)$ is $\sim$3.6$\times 10^{6}$ cm$^{-1}$ at $\sim$1.7 eV in the near-visible light region from the BSE calculation. The lowest reflectivity $r(\omega)$ is predicted to be around 2.3 \% in the active region of the solar energy spectrum, suggesting potential suitability for anti-reflective coatings in solar cell applications. To better understand the excitons, we have examined their characteristics in reciprocal and real space. The reciprocal space distribution of the main bound bright exciton shows localized behavior. In contrast, the real-space distribution shows its delocalized behavior across all the unit cells, indicating a significant spatial extent. Additionally, using the $\alpha(\omega)$ spectrum obtained from the BSE method, the SLME has been estimated to be $\sim$15.4\% for a single-junction PV absorber with a film thickness of $\sim$0.5~$\mu$m. Finally, for the stability analysis of this compound, the phonon dispersion and thermodynamic stability have been calculated. Using the thermodynamic formalism of chemical potentials, $V_{Cl}$ and $V_{Tl}$ are identified as the most stable neutral point defects under Cl-poor and Cl-rich conditions, respectively. Thus, tuning the chlorine chemical potential during synthesis can effectively control defect formation in RbTlCl$_3$. These results suggest that RbTlCl$_3$ is a promising candidate for potential energy application, although further experimental validation and device-level investigations are required.

\vspace{0.2in}
\section*{Acknowledgement}
We acknowledge the computational support provided by the High-Performance Computing (HPC) PARAM Himalaya at the Indian Institute of Technology Mandi.
	
	

\bibliography{ref.bib}

\end{document}